\documentclass[prb,twocolumn,superscriptaddress,longbibliography]{revtex4-2}
\usepackage{amsmath}
\usepackage{amssymb}
\usepackage{color}
\usepackage{graphicx}
\usepackage[colorlinks,bookmarks=false,citecolor=blue,linkcolor=red,urlcolor=blue]{hyperref}
\usepackage{booktabs}

\usepackage{cancel}     
\usepackage{soul}  

\begin{document}

\title{Spontaneous continuous-symmetry breaking and tower of states in a comb chain}

\author{Jingya Wang}
\affiliation{State Key Laboratory of Surface Physics and Department of Physics, Fudan University, Shanghai 200438, China}
\affiliation{Department of Physics, School of Science and Research Center for Industries of the Future, Westlake University, Hangzhou 310030,  China}
\affiliation{Institute of Natural Sciences, Westlake Institute for Advanced Study, Hangzhou 310024, China}

\author{Zenan Liu}
\affiliation{Department of Physics, School of Science and Research Center for Industries of the Future, Westlake University, Hangzhou 310030,  China}
\affiliation{Institute of Natural Sciences, Westlake Institute for Advanced Study, Hangzhou 310024, China}

\author{Bin-Bin Mao}
\affiliation{School of Foundational Education, University of Health and Rehabilitation Sciences, Qingdao 266000, China}

\author{Xu Tian}
\affiliation{Department of Physics, School of Science and Research Center for Industries of the Future, Westlake University, Hangzhou 310030,  China}
\affiliation{Institute of Natural Sciences, Westlake Institute for Advanced Study, Hangzhou 310024, China}

\author{Zijian Xiong}
\email{xiongzjsysu@hotmail.com}
\affiliation{College of Physics and Electronic Engineering, Chongqing Normal University, Chongqing 401331, China}

\author{Zhe Wang}
\email{wangzhe90@westlake.edu.cn}
\affiliation{Department of Physics, School of Science and Research Center for Industries of the Future, Westlake University, Hangzhou 310030, China}
\affiliation{Institute of Natural Sciences, Westlake Institute for Advanced Study, Hangzhou 310024, China}

\author{Zheng Yan}
\affiliation{Department of Physics, School of Science and Research Center for Industries of the Future, Westlake University, Hangzhou 310030, China}
\affiliation{Institute of Natural Sciences, Westlake Institute for Advanced Study, Hangzhou 310024, China}

\begin{abstract}
Based on the study of a one-dimensional (1D) antiferromagnetic Heisenberg model on a comb lattice, this work identifies an example of spontaneous continuous symmetry breaking in a 1D system with short-range interactions. When a symmetry-preserving relevant perturbation is applied to the system, we find that this model can always be described by the Marshall-Lieb-Mattis (MLM) theorem. The Shen-Qiu-Tian theorem establishes a direct connection between the MLM theorem (in the case of bipartite lattices with unequal numbers of sites in the two sublattices) and the breaking of continuous symmetry. 
Moreover, although authors of previous studies have suggested that the presence of a tower of states (TOS) serves as an important numerical diagnostic of the tendency of a system toward spontaneous symmetry breaking, these investigations have primarily focused on two-dimensional systems. In 1D systems, however, the presence of long-range order does not automatically imply the emergence of a TOS. Here, we observe the existence of a TOS in a 1D realistic ferrimagnetic lattice system with short-range interactions.

\end{abstract}

\date{\today}
\maketitle

\section{Introduction} 
One-dimensional (1D) quantum systems are rather special and deserve a separate introduction. 
Although the spectrum of the spin-$1/2$ Heisenberg chain was solved exactly by Bethe about 80 years ago~\cite{1Bethe1931}, low-dimensional quantum spin systems are still full of surprise and confusion. Lieb-Schultz-Mattis (LSM) theorem states that a 1D locally interacting half-integer spin chain with translation and spin-rotation symmetry cannot have a non-degenerate gapped ground state~\cite{Elliott1961Two,Affleck1986A}.The Haldane conjecture predicted that the Heisenberg
chain has completely different low-energy state properties with integer spin and half-integer spin~\cite{2Haldane1983,3Haldane1985,4Haldane1988}. The Hohenberg-Mermin-Wagner (HMW) theorem restricts the spontaneous breaking of continuous symmetries in a 1D quantum system~\cite{Mermin1966,Hohenberg1967,Sondhi1997Continuous,PhysRevB.109.L100502,PhysRevA.109.L011302,dljc-j3z7}.

The proof of the HMW theorem is mathematically rigorous, and thus, it dominates the understanding of symmetry breaking in 1D quantum systems~\cite{Mermin1966,Hohenberg1967,Sondhi1997Continuous}. Any violations of the HMW theorem are attributed to the unique low-energy state properties of the system, which are not generic. The 1D ferromagnetic Heisenberg chain and its variants can evade the constraints of this theorem because the order parameter is the generator of the global symmetry and commutes with the Hamiltonian, implying the absence of any quantum fluctuations~\cite{Anderson1994Basic,Zvonarev2007Spin,Aron2019An,Watanabe2020Counting}. 
Recent violations of HMW theorem in the (1+1)D surface of two-dimensional (2D) quantum systems have been attributed to the multi-mode coupling between the surface and the bulk~\cite{Zhang2017,Ding2018,Weber2018,Weber2019, Zhu2021, Wang2022,Wang2023}. Furthermore, Haruki Watanabe et al. mathematically constructed a class of quantum spin models that can realize spontaneous breaking of U(1) symmetry in 1D. More importantly, the order parameter does not commute with the Hamiltonian~\cite{Watanabe2024Critical}. The models share a common feature: Their Hamiltonians $H$ are frustration-free~\cite{tasaki2020physics}, meaning the ground state of $H=\sum_i^{L}H_i$ minimizes simultaneously all $H_i$, although $H_i$ do not need to commute with each other. 

In addition, Marshall-Lieb-Mattis (MLM) ferrimagnets, due to their nonlinear dispersion, fall outside the scope of the HMW theorem. Under the three conditions of a connected lattice (where any two sites can be linked by a series of bonds), a bipartite lattice, and antiferromagnetic (AFM) Heisenberg interactions, the MLM theorem  states that the ground state has a total spin of $S_{\text{tot}} = |N_1 - N_2| S$ and is $(2S_{\text{tot}} + 1)$-fold degenerate, where $N_1$ and $N_2$ are the number of spins on the two sublattices$^*$ (the distinction between sublattice$^*$ and sublattice is explained in Fig.~\ref{comb}) \cite{tasaki2020physics}. This theorem establishes that the ground state is degenerate. Thirty years later, under the same set of assumptions, the Shen–Qiu–Tian (SQT) theorem connected this result to the existence of long-range order, proving that the MLM theorem with $N_1 \neq N_2$ implies the existence of long-range ferrimagnetic order in the system \cite{shen1994ferrimagnetic,shen1998strongly}. Rigorously, these two theorems are based on the Perron-Frobenius theorem \cite{tasaki2020physics}, which is valid for some finite-dimensional matrices. 

In this paper, we study a spin-1/2 AFM Heisenberg model defined on a comb chain (as shown in Fig.~\ref{comb}). We find that the model indeed satisfies the conditions required by both the MLM theorem and the SQT theorem. We discover that, under these conditions, 
our numerical calculation results are consistent with the theorems.

Our innovation is the discovery of the existence of a tower of states (TOS) in a realistic 1D lattice system with short-range interactions. Although authors of previous studies \cite{lhuillier2002frustrated, laeuchli2016studying} have suggested that the presence of a TOS serves as an important numerical diagnostic of the tendency of a system toward spontaneous symmetry breaking (SSB),
these investigations have primarily focused on 2D systems. In 1D systems, however, the presence of long-range order does not automatically imply the emergence of a TOS. The SSB in the 1D ferromagnetic Heisenberg chain and its variants does not correspond to the emergence of a TOS~\cite{Anderson1994Basic,Zvonarev2007Spin,Aron2019An,Watanabe2020Counting,Watanabe2024Critical}.  In Ref.~\cite{Watanabe2024Critical}, a class of 1D models was proposed that exhibits spontaneous breaking of U(1) symmetry but lacks a TOS. More relevant to our work, 1D ferrimagnetic order does not always imply the presence of a TOS \cite{beekman2015criteria,rademaker2020stability}. Within our current understanding, no example of a TOS has been found in 1D lattice systems with short-range interactions. While the TOS is not a universal feature of all 1D symmetry‑broken phases, its presence in this model offers insight into the interplay between order, degeneracy, and low‑energy excitations in 1D.

\section{Model and method} 
\begin{figure}[htbp]
	\centering
	\includegraphics[width=0.35\textwidth]{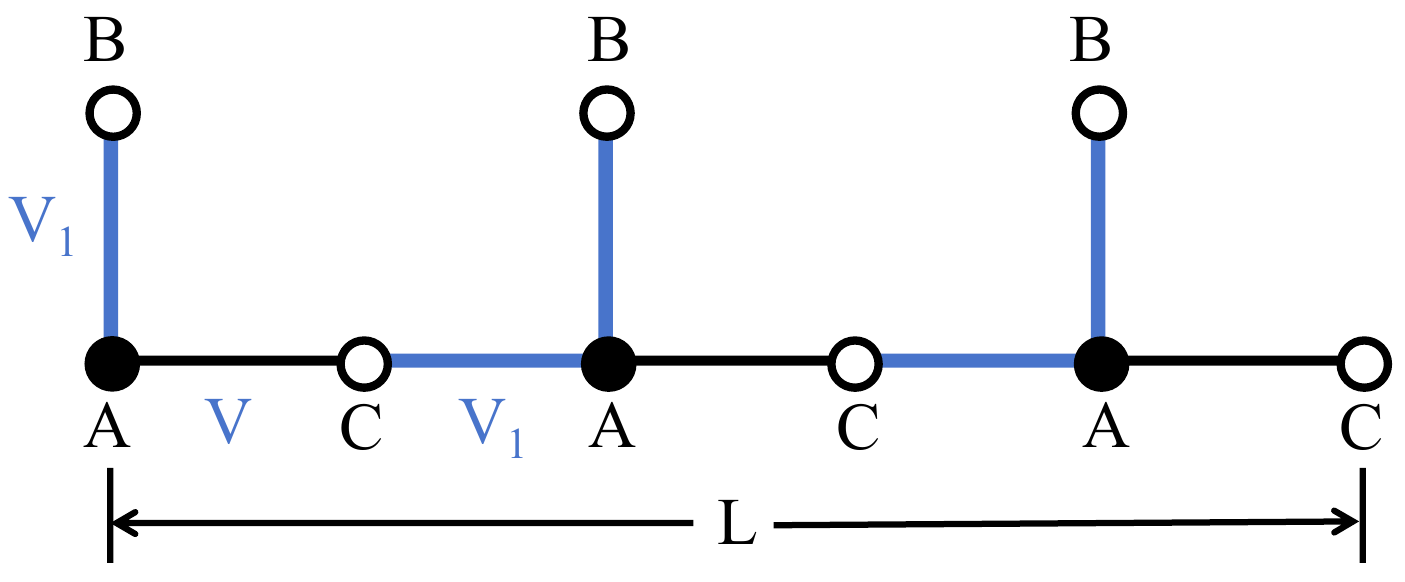}
	\caption{ A 1D comb lattice. $V = 1$ and $V_1 > 0$ represent AFM  Heisenberg interactions between nearest neighbors. This model is a bipartite lattice, where the sites of the lattice can be decomposed into two colors. In addition, from the perspective of lattice translation symmetry, this lattice can be divided into three sublattices: A, B, and C. For clarity, we use the term sublattice* when referring to the bipartite structure, and sublattice when discussing the translational lattice structure.}
    \label{comb}
\end{figure}

We consider a spin-1/2 AFM Heisenberg model defined on a comb chain with periodic boundary condition, see Fig.~\ref{comb}. For example, the Hamiltonian in Fig.~\ref{comb} can be written as
\begin{equation}\label{model}
H=\sum_{i}(V\textbf{S}_{i,A}\cdot\textbf{S}_{i,C}+V_{1}\textbf{S}_{i,A}\cdot\textbf{S}_{i,B}+V_1\textbf{S}_{i,C}\cdot \textbf{S}_{i+1,A}),
\end{equation}
where $i$ labels the unit cell, and each unit cell contains three spins (denoted by $A,B,$ and $C$). This comb chain is a bipartite lattice, where the sites of the lattice can be decomposed into two colors (black and white dots in Fig.~\ref{comb}), and any bond connects sites in different colors.

In this work, we use the stochastic series expansion (SSE) quantum Monte Carlo (QMC) algorithm \cite{Sandviksusc1991, Sandvik1999, yan2019sweeping, yan2022global} to explore the properties of the ground state of the system and demonstrate that, if the interactions are non-zero, the system is always described by the MLM ferrimagnetic and the SQT theorems. 
In our simulations, we have reached linear sizes up to $L=512$, and the inverse temperature scales as $\beta=2L$. Typically $10^{8}$ Monte Carlo samples are taken for each coupling strength.  In addition to the QMC calculations, we provide a reasonable explanation for the results by combining spin-wave theory and renormalization group theory.

\begin{figure*}[htbp]
	\centering 
	\includegraphics[width=1.0\textwidth]{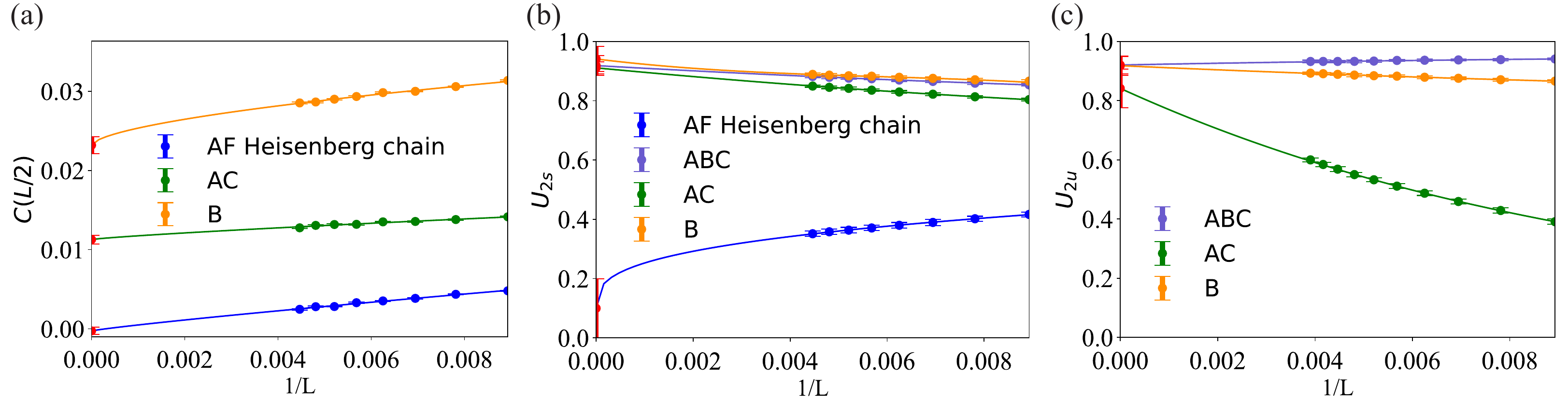}
	\caption{(a)–(c) Correlation $C(L/2)$, Binder cumulant $ U_2$ vs the inverse of the system size $1/L$ at $V=V_1=1$. $AC, B$ or $ABC$ indicate that calculated physical quantities belong to the $AC$ or $B$ sublattice or the whole system, as shown in Fig.~\ref{comb}. For comparison, we also computed the relevant physical results for a pure 1D AFM Heisenberg chain in (a) and (b). The red points indicate the extrapolated results in the thermodynamic limit.}
    \label{order}
\end{figure*}

\begin{figure}[htbp]
	\centering 
	\includegraphics[width=0.5\textwidth]{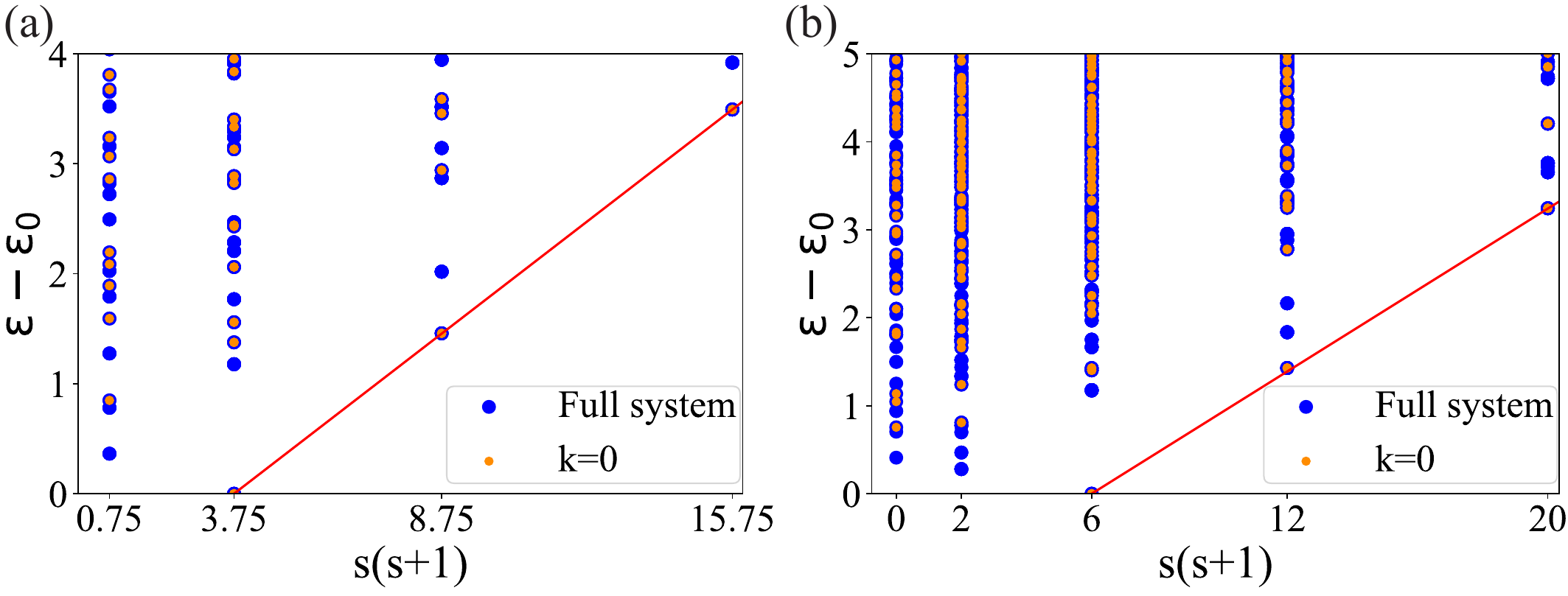}
	\caption{Energy spectra of a 1D comb lattice. $V = V_1 = 1$ with system sizes (a) $L = 6$  and (b) $L = 8$. 
    The blue dots represent the full energy spectrum, while the orange dots correspond to the spectrum extracted from the $k=0$ momentum subspace. The TOS levels are connected by red lines.  }
    \label{op}
\end{figure}

\section{MLM ferrimagnetic order}
We choose $ V = V_1 $ (as shown in Fig.~\ref{comb}) to study the properties of ferrimagnetic order.
First, we investigate the SSB
in the system using  QMC simulations.  We calculate the correlation $C(L/2)$ between two spins, either belonging to sublattice $B$ or $AC$ (see Fig.~\ref{comb}), separated by the longest distance $|i-j|=L/2$. 

The numerical results of correlations as functions of $L$  are shown in Fig.~\ref{order} (a). Both correlation functions tend to finite values in the thermodynamic limit. We try to fit the curves with
   $C(L/2) = c + aL^{-p}$
and find $c=0.023(1)$ for $BB$ correlation and  $c=0.0113(6)$ for the sublattice $AC$
as $L\to \infty$,  indicating the presence of long-range order. To facilitate comparison, we also calculated the correlation for the 1D AFM Heisenberg chain (the interaction between $A$ and $B$ is zero in Fig. \ref{comb}), which clearly tends to zero in the thermodynamic limit, as shown in Fig. \ref{order} (a).

 To further determine the long-range order of the  system, we calculate the Binder cumulant $U_{2s(u)}$ \cite{Binder1981, Binder1984}, which is defined based on staggered  magnetization $m_{s}(L)$ and uniform magnetization $m_{u}(L)$ as follows: 
\begin{equation}
 U_{2s(u)}(L)= \frac{5}{6}\left(3-\frac{\langle [m_{s(u)}^z(L)]^{4}\rangle}{\langle [m_{s(u)}^z(L)]^{2}\rangle^{2}}\right),
 \end{equation}
where $m_s^z=\frac{1}{N}\sum_i \phi_{i} S_i^z$, with the staggered phase factor $\phi_{i}=\pm 1$ depending on the sublattice$^*$ and $m_u^z=\frac{1}{N}\sum_i S_i^z$ and  $N$ is the number of spins included in the sum. In this work, we
consider three cases: $N$ belongs to the 1D comb lattice ($ABC$), sublattice $B$ or sublattice $AC$, as shown in Fig.~\ref{comb}. Here, $U_{2s(u)}(L)$ converges to 1 as $L \to \infty$, indicating the existence of magnetic order associated with SSB from O(3) to O(2), and approaches zero with increasing system size, implying that the system is in the magnetically disordered phase.

The numerical results of $ U_{2s}(L)$ for sublattice $B$ or sublattice $AC$ or the whole system $ABC$ as a function of size $1/L$ are plotted in Fig. \ref{order} (b). We fit the data using a polynomial in  $1/L$
as $U_{2s}(L) =U_{2s}+c_{1}L^{-1}+c_{2}L^{-2}+c_{3}L^{-3}$.
We find statistically sound estimation  $U_{2s}=0.92(3)$ for $ABC$, $U_{2s}=0.94(4)$ for sublattice $B$ and $U_{2}=0.91(2)$ for sublattice $AC$, all of which are close to 1 within the error bar. The numerical results of $ U_{2u}(L)$ for sublattice $B$, sublattice $AC$ or the whole system $ABC$ as a function of size $1/L$ are plotted in Fig. \ref{order} (c). We fit the data using the polynomial given above
 and find   $U_{2u}=0.92(2)$ for $ABC$, $U_{2u}=0.92(3)$ for sublattice $B$ and $U_{2u}=0.84(7)$ for sublattice $AC$, all of which are close to 1 within the error bar. This indicates that the ground state of the system simultaneously exhibits staggered magnetization and uniform magnetization.

\section{Tower of states}
Although the presence of a TOS~\cite{Anderson1994Basic} has been proposed as a signature of SSB,
its necessity has not been established, with most supporting evidence coming from 2D systems. As discussed in our Introduction, in 1D systems, long-range order does not guarantee the emergence of a TOS ~\cite{Anderson1994Basic,Zvonarev2007Spin,Aron2019An,Watanabe2020Counting,Watanabe2024Critical,beekman2015criteria,rademaker2020stability}. Detecting the presence or absence of a TOS in realistic 1D lattice models with short-range interactions remains an open question.

For the model studied in this paper, we find that the TOS clearly emerges in the momentum subspace at $k=0$, as shown in Figs.~\ref{op}(a) and \ref{op}(b) in the symmetry-broken phase.  
Since the comb chain is bipartite, the ground state wave function obeys the Marshall sign structure \cite{Schnack2000jmmm, tasaki2020physics}. On the other hand, translation does not exchange the two bipartite sublattices${^*}$. 
Hence, the sign structure remains invariant under translation, which implies that the ground state has momentum $k=0$. The energies of the TOS in finite size obey the relation
\begin{equation}
\epsilon-\epsilon_0\propto {S(S+n-2)},
\label{tos}
\end{equation}
where $S$ is the total spin momentum of the system, $n$ means that the ground state breaks an $O(n)$ symmetry (in this work, $n=3$)~\cite{lhuillier2002frustrated,wietek2017studying,mao2025sampling,mao2025arxiv}. As shown in Figs.~\ref{op}(a) and \ref{op}(b), the TOS levels are connected by red lines.

Moreover, we demonstrate numerically that our model is described by the MLM theorem. The MLM theorem for bipartite lattices states that the ground state has a total spin of $S_{\text{tot}} = |N_1 - N_2| S$ and is $(2S_{\text{tot}} + 1)$-fold degenerate, where $N_1$ and $N_2$ are the number of spins on the two sublattices$^*$ \cite{tasaki2020physics}.  In our system ($S = 1/2$, with $A$ belonging to one sublattice$^*$ and $BC$ to the other), for $L = 6$, the ground state has $S_{\text{tot}} = 3/2$, leading to $S_{\text{tot}}(S_{\text{tot}} + 1) = 3.75$; for $L = 8$, the ground state has $S_{\text{tot}} = 2$, resulting in $S_{\text{tot}}(S_{\text{tot}} + 1) = 6$. As shown in Fig.~\ref{op}, our numerical results are in full agreement with the predictions of the theorem. We have also verified the ground-state degeneracy, which is consistent with theoretical expectations.

\begin{figure}[htbp]
	\centering
	\includegraphics[width=0.3\textwidth]{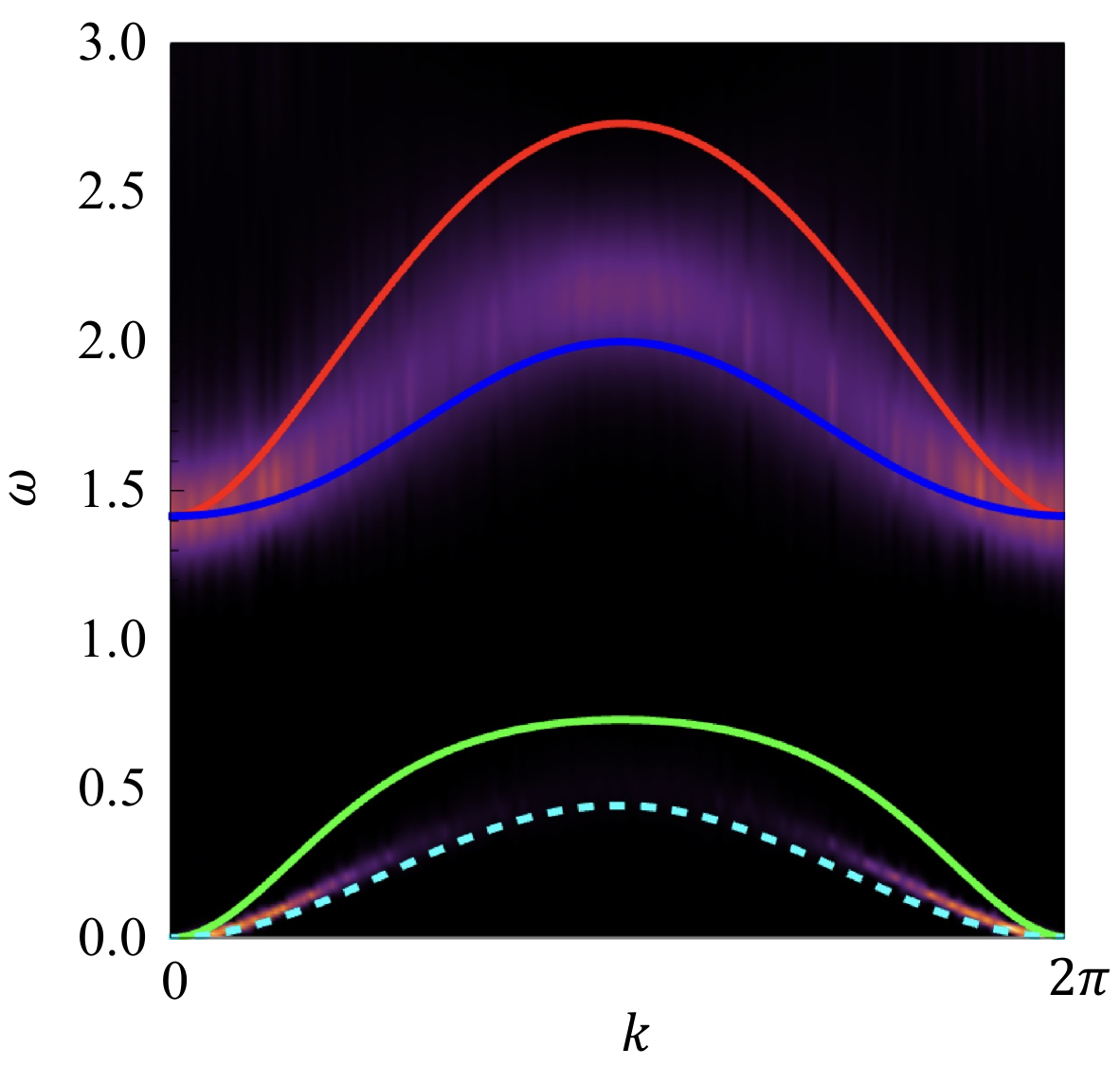}
	\caption{\label{rob,spinwavedis}  Dispersions calculated by linear spin wave theory (green, blue and red solid lines), low-energy effective model from Kadanoff's approach (dashed line) and SAC (spectral functions marked by color intensity) of the comb chain, where $V_1=V=1$.}
\end{figure}

\section{Spin-wave dispersions and Low-energy effective theory}
As discussed above, the model defined by Eq. (1) falls within the scope of the SQT theorem, which guarantees well-defined long-range ferrimagnetic order for any $V_1 > 0$. Hence, the excitations are spin waves in this 
ferrimagnetic order. The results calculated by spin wave theory (see Appendix \ref{appendixd}) and QMC-stochastic analytic continuation (SAC)~\cite{sandvik2016constrained,shao2017nearly,shao2023progress} are shown in \textcolor{blue}{Fig.~\ref{rob,spinwavedis}}. The SAC method can extract the spectrum [dynamical structure factor $S(\bf{k},\omega)$] from the QMC data of imaginary time correlations. The details of the SAC calculations are provided in Appendix \ref{appendixb}. 

In Fig.~\ref{rob,spinwavedis}, we find that there are gapless low-energy ($\omega<0.5$) and high-energy continuum ($\omega>1$) branches. In addition, the low-energy branch is quadratic around $k\sim 0$ and the spin wave theory (green, blue, and red lines in Fig.~\ref{rob,spinwavedis}) qualitatively agrees with the spectrum obtained from the QMC-SAC result. 
All these features are consistent with the double-branch excitation spectrum of the Heisenberg ferrimagnetic chain, which is characterized by a low-energy gapless ferromagnetic-like branch and a high-energy gapped AFM-like branch \cite{PhysRevB.60.1057}.

We have checked that within a certain range for $V_1$, the quadratic behavior $\omega \sim k^2$ remains, which further supports the robustness of the SSB mentioned above. Moreover, the high-energy continuum obtained from QMC-SAC appears to be bounded by the upper two branches of the dispersions from spin wave theory. This continuum may imply that there are interactions among the high-energy spin waves, but we are only interested in the low-energy behavior in this work.

To gain a better understanding for the low-energy quadratic behavior, we use Kadanoff's renormalization group approach \cite{Langari1998prb,Langari2008pra} (see Appendix \ref{appendixe}) and show that the low-energy effective theory of the AFM model Eq.~(\ref{model}) is actually a ferromagnetic Heisenberg chain with dispersion $2SJ_{eff}(1-\cos k)$ \cite{karbach1997}, and $J_{eff}=2/9$ for $V=V_1=1$. This effective model explains the quadratic dispersion and matches the QMC-SAC spectrum nicely (dashed line in Fig.~\ref{rob,spinwavedis}). Furthermore, it is also known that continuous symmetry breaking with a quadratic gapless mode in 1D does not violate the HMW theorem \cite{Watanabe2024Critical,Watanabe2020Counting,Aron2019An} (see Appendix \ref{appendixf}). This is consistent with the TOS results.

\section{Conclusion} 
In this work, we investigate the spin-1/2 AFM Heisenberg model defined on a comb-like chain. We identify the presence of a TOS in this 1D lattice system with short-range interactions. Although previous studies have pointed out that the TOS serves as an important numerical diagnostic for detecting tendencies toward SSB,
most of those works have focused on 2D systems. It has been shown that 1D systems with short-range interactions generally lack a TOS structure. We confirm that this model satisfies the total spin and degeneracy of the ground state predicted by the MLM theorem, as well as the long-range order predicted by the SQT theorem. This finding provides a perspective for understanding the relationship between symmetry breaking and the structure of low-energy excitations in 1D systems.

\section{acknowledgments}
We are grateful for the helpful discussions with Haruki Watanabe, Shun-Qing Shen, Chunhao Guo and Hosho Katsura. 
Z. X. acknowledges funding from
the National Science Foundation of China (Grants No. 12404169 ), and start-up funding of Chongqing
Normal University (Grant No. 24XLB010).
ZW is supported by the China Postdoctoral Science Foundation under Grant No.2024M752898. 
Z.L. is supported by the China Postdoctoral Science
Foundation under Grant No.2024M762935.
This project is supported by the Scientific Research Fund for Distinguished Young Scholars of the Education Department of Anhui Province (No.2022AH020008), the Scientific Research Project
(No.WU2025B011) 
and the Start-up Funding of Westlake University. The authors thank the high-performance computing centers of Westlake University and the Beijing PARATERA Tech Co., Ltd. for providing HPC resources.

\emph{Note added.}  While we were compiling this manuscript we became aware of a preprint \cite{Nahum2025} that proposes the spontaneous breaking of the U(1) symmetry at the phase transition by perturbing the 1D Heisenberg ferromagnet. 
 
\emph{Data availability.---}  The data that support the findings of this article are openly available \cite{wang}

\bibliographystyle{apsrev4-2}
\bibliography{ref}

\begin{thebibliography}{54}%
\makeatletter
\providecommand \@ifxundefined [1]{%
 \@ifx{#1\undefined}
}%
\providecommand \@ifnum [1]{%
 \ifnum #1\expandafter \@firstoftwo
 \else \expandafter \@secondoftwo
 \fi
}%
\providecommand \@ifx [1]{%
 \ifx #1\expandafter \@firstoftwo
 \else \expandafter \@secondoftwo
 \fi
}%
\providecommand \natexlab [1]{#1}%
\providecommand \enquote  [1]{``#1''}%
\providecommand \bibnamefont  [1]{#1}%
\providecommand \bibfnamefont [1]{#1}%
\providecommand \citenamefont [1]{#1}%
\providecommand \href@noop [0]{\@secondoftwo}%
\providecommand \href [0]{\begingroup \@sanitize@url \@href}%
\providecommand \@href[1]{\@@startlink{#1}\@@href}%
\providecommand \@@href[1]{\endgroup#1\@@endlink}%
\providecommand \@sanitize@url [0]{\catcode `\\12\catcode `\$12\catcode `\&12\catcode `\#12\catcode `\^12\catcode `\_12\catcode `\%12\relax}%
\providecommand \@@startlink[1]{}%
\providecommand \@@endlink[0]{}%
\providecommand \url  [0]{\begingroup\@sanitize@url \@url }%
\providecommand \@url [1]{\endgroup\@href {#1}{\urlprefix }}%
\providecommand \urlprefix  [0]{URL }%
\providecommand \Eprint [0]{\href }%
\providecommand \doibase [0]{https://doi.org/}%
\providecommand \selectlanguage [0]{\@gobble}%
\providecommand \bibinfo  [0]{\@secondoftwo}%
\providecommand \bibfield  [0]{\@secondoftwo}%
\providecommand \translation [1]{[#1]}%
\providecommand \BibitemOpen [0]{}%
\providecommand \bibitemStop [0]{}%
\providecommand \bibitemNoStop [0]{.\EOS\space}%
\providecommand \EOS [0]{\spacefactor3000\relax}%
\providecommand \BibitemShut  [1]{\csname bibitem#1\endcsname}%
\let\auto@bib@innerbib\@empty
\bibitem [{\citenamefont {Bethe}(1931)}]{1Bethe1931}%
  \BibitemOpen
  \bibfield  {author} {\bibinfo {author} {\bibfnamefont {H.}~\bibnamefont {Bethe}},\ }\href {https://doi.org/10.1007/BF01341708} {\bibfield  {journal} {\bibinfo  {journal} {Zeitschrift f{\"u}r Physik}\ }\textbf {\bibinfo {volume} {71}},\ \bibinfo {pages} {205} (\bibinfo {year} {1931})}\BibitemShut {NoStop}%
\bibitem [{\citenamefont {Lieb}\ \emph {et~al.}(1961)\citenamefont {Lieb}, \citenamefont {Schultz},\ and\ \citenamefont {Mattis}}]{Elliott1961Two}%
  \BibitemOpen
  \bibfield  {author} {\bibinfo {author} {\bibfnamefont {E.}~\bibnamefont {Lieb}}, \bibinfo {author} {\bibfnamefont {T.}~\bibnamefont {Schultz}},\ and\ \bibinfo {author} {\bibfnamefont {D.}~\bibnamefont {Mattis}},\ }\href {https://doi.org/https://doi.org/10.1016/0003-4916(61)90115-4} {\bibfield  {journal} {\bibinfo  {journal} {Annals of Physics}\ }\textbf {\bibinfo {volume} {16}},\ \bibinfo {pages} {407} (\bibinfo {year} {1961})}\BibitemShut {NoStop}%
\bibitem [{\citenamefont {Affleck}\ and\ \citenamefont {Lieb}(1986)}]{Affleck1986A}%
  \BibitemOpen
  \bibfield  {author} {\bibinfo {author} {\bibfnamefont {I.}~\bibnamefont {Affleck}}\ and\ \bibinfo {author} {\bibfnamefont {E.~H.}\ \bibnamefont {Lieb}},\ }\href {https://doi.org/10.1007/BF00400304} {\bibfield  {journal} {\bibinfo  {journal} {Letters in Mathematical Physics}\ }\textbf {\bibinfo {volume} {12}},\ \bibinfo {pages} {57} (\bibinfo {year} {1986})}\BibitemShut {NoStop}%
\bibitem [{\citenamefont {Haldane}(1983)}]{2Haldane1983}%
  \BibitemOpen
  \bibfield  {author} {\bibinfo {author} {\bibfnamefont {F.~D.~M.}\ \bibnamefont {Haldane}},\ }\href {https://doi.org/10.1103/PhysRevLett.50.1153} {\bibfield  {journal} {\bibinfo  {journal} {Phys. Rev. Lett.}\ }\textbf {\bibinfo {volume} {50}},\ \bibinfo {pages} {1153} (\bibinfo {year} {1983})}\BibitemShut {NoStop}%
\bibitem [{\citenamefont {Haldane}(1985)}]{3Haldane1985}%
  \BibitemOpen
  \bibfield  {author} {\bibinfo {author} {\bibfnamefont {F.~D.~M.}\ \bibnamefont {Haldane}},\ }\href {https://doi.org/10.1063/1.335096} {\bibfield  {journal} {\bibinfo  {journal} {Journal of Applied Physics}\ }\textbf {\bibinfo {volume} {57}},\ \bibinfo {pages} {3359} (\bibinfo {year} {1985})}\BibitemShut {NoStop}%
\bibitem [{\citenamefont {Haldane}(1988)}]{4Haldane1988}%
  \BibitemOpen
  \bibfield  {author} {\bibinfo {author} {\bibfnamefont {F.~D.~M.}\ \bibnamefont {Haldane}},\ }\href {https://doi.org/10.1103/PhysRevLett.61.1029} {\bibfield  {journal} {\bibinfo  {journal} {Phys. Rev. Lett.}\ }\textbf {\bibinfo {volume} {61}},\ \bibinfo {pages} {1029} (\bibinfo {year} {1988})}\BibitemShut {NoStop}%
\bibitem [{\citenamefont {Mermin}\ and\ \citenamefont {Wagner}(1966)}]{Mermin1966}%
  \BibitemOpen
  \bibfield  {author} {\bibinfo {author} {\bibfnamefont {N.~D.}\ \bibnamefont {Mermin}}\ and\ \bibinfo {author} {\bibfnamefont {H.}~\bibnamefont {Wagner}},\ }\href {https://doi.org/10.1103/PhysRevLett.17.1133} {\bibfield  {journal} {\bibinfo  {journal} {Phys. Rev. Lett.}\ }\textbf {\bibinfo {volume} {17}},\ \bibinfo {pages} {1133} (\bibinfo {year} {1966})}\BibitemShut {NoStop}%
\bibitem [{\citenamefont {Hohenberg}(1967)}]{Hohenberg1967}%
  \BibitemOpen
  \bibfield  {author} {\bibinfo {author} {\bibfnamefont {P.~C.}\ \bibnamefont {Hohenberg}},\ }\href {https://doi.org/10.1103/PhysRev.158.383} {\bibfield  {journal} {\bibinfo  {journal} {Phys. Rev.}\ }\textbf {\bibinfo {volume} {158}},\ \bibinfo {pages} {383} (\bibinfo {year} {1967})}\BibitemShut {NoStop}%
\bibitem [{\citenamefont {Sondhi}\ \emph {et~al.}(1997)\citenamefont {Sondhi}, \citenamefont {Girvin}, \citenamefont {Carini},\ and\ \citenamefont {Shahar}}]{Sondhi1997Continuous}%
  \BibitemOpen
  \bibfield  {author} {\bibinfo {author} {\bibfnamefont {S.~L.}\ \bibnamefont {Sondhi}}, \bibinfo {author} {\bibfnamefont {S.~M.}\ \bibnamefont {Girvin}}, \bibinfo {author} {\bibfnamefont {J.~P.}\ \bibnamefont {Carini}},\ and\ \bibinfo {author} {\bibfnamefont {D.}~\bibnamefont {Shahar}},\ }\href {https://doi.org/10.1103/RevModPhys.69.315} {\bibfield  {journal} {\bibinfo  {journal} {Rev. Mod. Phys.}\ }\textbf {\bibinfo {volume} {69}},\ \bibinfo {pages} {315} (\bibinfo {year} {1997})}\BibitemShut {NoStop}%
\bibitem [{\citenamefont {Kuklov}\ \emph {et~al.}(2024{\natexlab{a}})\citenamefont {Kuklov}, \citenamefont {Prokof'ev}, \citenamefont {Radzihovsky},\ and\ \citenamefont {Svistunov}}]{PhysRevB.109.L100502}%
  \BibitemOpen
  \bibfield  {author} {\bibinfo {author} {\bibfnamefont {A.}~\bibnamefont {Kuklov}}, \bibinfo {author} {\bibfnamefont {N.}~\bibnamefont {Prokof'ev}}, \bibinfo {author} {\bibfnamefont {L.}~\bibnamefont {Radzihovsky}},\ and\ \bibinfo {author} {\bibfnamefont {B.}~\bibnamefont {Svistunov}},\ }\href {https://doi.org/10.1103/PhysRevB.109.L100502} {\bibfield  {journal} {\bibinfo  {journal} {Phys. Rev. B}\ }\textbf {\bibinfo {volume} {109}},\ \bibinfo {pages} {L100502} (\bibinfo {year} {2024}{\natexlab{a}})}\BibitemShut {NoStop}%
\bibitem [{\citenamefont {Kuklov}\ \emph {et~al.}(2024{\natexlab{b}})\citenamefont {Kuklov}, \citenamefont {Pollet}, \citenamefont {Prokof'ev}, \citenamefont {Radzihovsky},\ and\ \citenamefont {Svistunov}}]{PhysRevA.109.L011302}%
  \BibitemOpen
  \bibfield  {author} {\bibinfo {author} {\bibfnamefont {A.}~\bibnamefont {Kuklov}}, \bibinfo {author} {\bibfnamefont {L.}~\bibnamefont {Pollet}}, \bibinfo {author} {\bibfnamefont {N.}~\bibnamefont {Prokof'ev}}, \bibinfo {author} {\bibfnamefont {L.}~\bibnamefont {Radzihovsky}},\ and\ \bibinfo {author} {\bibfnamefont {B.}~\bibnamefont {Svistunov}},\ }\href {https://doi.org/10.1103/PhysRevA.109.L011302} {\bibfield  {journal} {\bibinfo  {journal} {Phys. Rev. A}\ }\textbf {\bibinfo {volume} {109}},\ \bibinfo {pages} {L011302} (\bibinfo {year} {2024}{\natexlab{b}})}\BibitemShut {NoStop}%
\bibitem [{\citenamefont {Radzihovsky}\ and\ \citenamefont {Pellett}(2025)}]{dljc-j3z7}%
  \BibitemOpen
  \bibfield  {author} {\bibinfo {author} {\bibfnamefont {L.}~\bibnamefont {Radzihovsky}}\ and\ \bibinfo {author} {\bibfnamefont {E.}~\bibnamefont {Pellett}},\ }\href {https://doi.org/10.1103/dljc-j3z7} {\bibfield  {journal} {\bibinfo  {journal} {Phys. Rev. Lett.}\ }\textbf {\bibinfo {volume} {135}},\ \bibinfo {pages} {016001} (\bibinfo {year} {2025})}\BibitemShut {NoStop}%
\bibitem [{\citenamefont {Anderson}(1994)}]{Anderson1994Basic}%
  \BibitemOpen
  \bibfield  {author} {\bibinfo {author} {\bibfnamefont {P.~W.}\ \bibnamefont {Anderson}},\ }\href {https://doi.org/10.4324/9780429494116} {\emph {\bibinfo {title} {Basic Notions of Condensed Matter Physics}}},\ \bibinfo {edition} {1st}\ ed.\ (\bibinfo  {publisher} {CRC Press},\ \bibinfo {year} {1994})\BibitemShut {NoStop}%
\bibitem [{\citenamefont {Zvonarev}\ \emph {et~al.}(2007)\citenamefont {Zvonarev}, \citenamefont {Cheianov},\ and\ \citenamefont {Giamarchi}}]{Zvonarev2007Spin}%
  \BibitemOpen
  \bibfield  {author} {\bibinfo {author} {\bibfnamefont {M.~B.}\ \bibnamefont {Zvonarev}}, \bibinfo {author} {\bibfnamefont {V.~V.}\ \bibnamefont {Cheianov}},\ and\ \bibinfo {author} {\bibfnamefont {T.}~\bibnamefont {Giamarchi}},\ }\href {https://doi.org/10.1103/PhysRevLett.99.240404} {\bibfield  {journal} {\bibinfo  {journal} {Phys. Rev. Lett.}\ }\textbf {\bibinfo {volume} {99}},\ \bibinfo {pages} {240404} (\bibinfo {year} {2007})}\BibitemShut {NoStop}%
\bibitem [{\citenamefont {Beekman}\ \emph {et~al.}(2019)\citenamefont {Beekman}, \citenamefont {Rademaker},\ and\ \citenamefont {van Wezel}}]{Aron2019An}%
  \BibitemOpen
  \bibfield  {author} {\bibinfo {author} {\bibfnamefont {A.~J.}\ \bibnamefont {Beekman}}, \bibinfo {author} {\bibfnamefont {L.}~\bibnamefont {Rademaker}},\ and\ \bibinfo {author} {\bibfnamefont {J.}~\bibnamefont {van Wezel}},\ }\href {https://doi.org/10.21468/SciPostPhysLectNotes.11} {\bibfield  {journal} {\bibinfo  {journal} {SciPost Phys. Lect. Notes}\ ,\ \bibinfo {pages} {11}} (\bibinfo {year} {2019})}\BibitemShut {NoStop}%
\bibitem [{\citenamefont {Watanabe}(2020)}]{Watanabe2020Counting}%
  \BibitemOpen
  \bibfield  {author} {\bibinfo {author} {\bibfnamefont {H.}~\bibnamefont {Watanabe}},\ }\href {https://doi.org/https://doi.org/10.1146/annurev-conmatphys-031119-050644} {\bibfield  {journal} {\bibinfo  {journal} {Annual Review of Condensed Matter Physics}\ }\textbf {\bibinfo {volume} {11}},\ \bibinfo {pages} {169} (\bibinfo {year} {2020})}\BibitemShut {NoStop}%
\bibitem [{\citenamefont {Zhang}\ and\ \citenamefont {Wang}(2017)}]{Zhang2017}%
  \BibitemOpen
  \bibfield  {author} {\bibinfo {author} {\bibfnamefont {L.}~\bibnamefont {Zhang}}\ and\ \bibinfo {author} {\bibfnamefont {F.}~\bibnamefont {Wang}},\ }\href {https://doi.org/10.1103/PhysRevLett.118.087201} {\bibfield  {journal} {\bibinfo  {journal} {Phys. Rev. Lett.}\ }\textbf {\bibinfo {volume} {118}},\ \bibinfo {pages} {087201} (\bibinfo {year} {2017})}\BibitemShut {NoStop}%
\bibitem [{\citenamefont {Ding}\ \emph {et~al.}(2018)\citenamefont {Ding}, \citenamefont {Zhang},\ and\ \citenamefont {Guo}}]{Ding2018}%
  \BibitemOpen
  \bibfield  {author} {\bibinfo {author} {\bibfnamefont {C.}~\bibnamefont {Ding}}, \bibinfo {author} {\bibfnamefont {L.}~\bibnamefont {Zhang}},\ and\ \bibinfo {author} {\bibfnamefont {W.}~\bibnamefont {Guo}},\ }\href {https://doi.org/10.1103/PhysRevLett.120.235701} {\bibfield  {journal} {\bibinfo  {journal} {Phys. Rev. Lett.}\ }\textbf {\bibinfo {volume} {120}},\ \bibinfo {pages} {235701} (\bibinfo {year} {2018})}\BibitemShut {NoStop}%
\bibitem [{\citenamefont {Weber}\ \emph {et~al.}(2018)\citenamefont {Weber}, \citenamefont {Parisen~Toldin},\ and\ \citenamefont {Wessel}}]{Weber2018}%
  \BibitemOpen
  \bibfield  {author} {\bibinfo {author} {\bibfnamefont {L.}~\bibnamefont {Weber}}, \bibinfo {author} {\bibfnamefont {F.}~\bibnamefont {Parisen~Toldin}},\ and\ \bibinfo {author} {\bibfnamefont {S.}~\bibnamefont {Wessel}},\ }\href {https://doi.org/10.1103/PhysRevB.98.140403} {\bibfield  {journal} {\bibinfo  {journal} {Phys. Rev. B}\ }\textbf {\bibinfo {volume} {98}},\ \bibinfo {pages} {140403} (\bibinfo {year} {2018})}\BibitemShut {NoStop}%
\bibitem [{\citenamefont {Weber}\ and\ \citenamefont {Wessel}(2019)}]{Weber2019}%
  \BibitemOpen
  \bibfield  {author} {\bibinfo {author} {\bibfnamefont {L.}~\bibnamefont {Weber}}\ and\ \bibinfo {author} {\bibfnamefont {S.}~\bibnamefont {Wessel}},\ }\href {https://doi.org/10.1103/PhysRevB.100.054437} {\bibfield  {journal} {\bibinfo  {journal} {Phys. Rev. B}\ }\textbf {\bibinfo {volume} {100}},\ \bibinfo {pages} {054437} (\bibinfo {year} {2019})}\BibitemShut {NoStop}%
\bibitem [{\citenamefont {Zhu}\ \emph {et~al.}(2021)\citenamefont {Zhu}, \citenamefont {Ding}, \citenamefont {Zhang},\ and\ \citenamefont {Guo}}]{Zhu2021}%
  \BibitemOpen
  \bibfield  {author} {\bibinfo {author} {\bibfnamefont {W.}~\bibnamefont {Zhu}}, \bibinfo {author} {\bibfnamefont {C.}~\bibnamefont {Ding}}, \bibinfo {author} {\bibfnamefont {L.}~\bibnamefont {Zhang}},\ and\ \bibinfo {author} {\bibfnamefont {W.}~\bibnamefont {Guo}},\ }\href {https://doi.org/10.1103/PhysRevB.103.024412} {\bibfield  {journal} {\bibinfo  {journal} {Phys. Rev. B}\ }\textbf {\bibinfo {volume} {103}},\ \bibinfo {pages} {024412} (\bibinfo {year} {2021})}\BibitemShut {NoStop}%
\bibitem [{\citenamefont {Wang}\ \emph {et~al.}(2022)\citenamefont {Wang}, \citenamefont {Zhang},\ and\ \citenamefont {Guo}}]{Wang2022}%
  \BibitemOpen
  \bibfield  {author} {\bibinfo {author} {\bibfnamefont {Z.}~\bibnamefont {Wang}}, \bibinfo {author} {\bibfnamefont {F.}~\bibnamefont {Zhang}},\ and\ \bibinfo {author} {\bibfnamefont {W.}~\bibnamefont {Guo}},\ }\href {https://doi.org/10.1103/PhysRevB.106.134407} {\bibfield  {journal} {\bibinfo  {journal} {Phys. Rev. B}\ }\textbf {\bibinfo {volume} {106}},\ \bibinfo {pages} {134407} (\bibinfo {year} {2022})}\BibitemShut {NoStop}%
\bibitem [{\citenamefont {Wang}\ \emph {et~al.}(2023)\citenamefont {Wang}, \citenamefont {Zhang},\ and\ \citenamefont {Guo}}]{Wang2023}%
  \BibitemOpen
  \bibfield  {author} {\bibinfo {author} {\bibfnamefont {Z.}~\bibnamefont {Wang}}, \bibinfo {author} {\bibfnamefont {F.}~\bibnamefont {Zhang}},\ and\ \bibinfo {author} {\bibfnamefont {W.}~\bibnamefont {Guo}},\ }\href {https://doi.org/10.1103/PhysRevB.108.014409} {\bibfield  {journal} {\bibinfo  {journal} {Phys. Rev. B}\ }\textbf {\bibinfo {volume} {108}},\ \bibinfo {pages} {014409} (\bibinfo {year} {2023})}\BibitemShut {NoStop}%
\bibitem [{\citenamefont {Watanabe}\ \emph {et~al.}(2024)\citenamefont {Watanabe}, \citenamefont {Katsura},\ and\ \citenamefont {Lee}}]{Watanabe2024Critical}%
  \BibitemOpen
  \bibfield  {author} {\bibinfo {author} {\bibfnamefont {H.}~\bibnamefont {Watanabe}}, \bibinfo {author} {\bibfnamefont {H.}~\bibnamefont {Katsura}},\ and\ \bibinfo {author} {\bibfnamefont {J.~Y.}\ \bibnamefont {Lee}},\ }\href {https://doi.org/10.1103/PhysRevLett.133.176001} {\bibfield  {journal} {\bibinfo  {journal} {Phys. Rev. Lett.}\ }\textbf {\bibinfo {volume} {133}},\ \bibinfo {pages} {176001} (\bibinfo {year} {2024})}\BibitemShut {NoStop}%
\bibitem [{\citenamefont {Tasaki}(2020)}]{tasaki2020physics}%
  \BibitemOpen
  \bibfield  {author} {\bibinfo {author} {\bibfnamefont {H.}~\bibnamefont {Tasaki}},\ }\href@noop {} {\emph {\bibinfo {title} {Physics and mathematics of quantum many-body systems}}},\ Vol.~\bibinfo {volume} {66}\ (\bibinfo  {publisher} {Springer},\ \bibinfo {year} {2020})\BibitemShut {NoStop}%
\bibitem [{\citenamefont {Shen}\ \emph {et~al.}(1994)\citenamefont {Shen}, \citenamefont {Qiu},\ and\ \citenamefont {Tian}}]{shen1994ferrimagnetic}%
  \BibitemOpen
  \bibfield  {author} {\bibinfo {author} {\bibfnamefont {S.-Q.}\ \bibnamefont {Shen}}, \bibinfo {author} {\bibfnamefont {Z.-M.}\ \bibnamefont {Qiu}},\ and\ \bibinfo {author} {\bibfnamefont {G.-S.}\ \bibnamefont {Tian}},\ }\href {https://doi.org/10.1103/PhysRevLett.72.1280} {\bibfield  {journal} {\bibinfo  {journal} {Phys. Rev. Lett.}\ }\textbf {\bibinfo {volume} {72}},\ \bibinfo {pages} {1280} (\bibinfo {year} {1994})}\BibitemShut {NoStop}%
\bibitem [{\citenamefont {Shen}(1998)}]{shen1998strongly}%
  \BibitemOpen
  \bibfield  {author} {\bibinfo {author} {\bibfnamefont {S.-Q.}\ \bibnamefont {Shen}},\ }\href@noop {} {\bibfield  {journal} {\bibinfo  {journal} {International Journal of Modern Physics B}\ }\textbf {\bibinfo {volume} {12}},\ \bibinfo {pages} {709} (\bibinfo {year} {1998})}\BibitemShut {NoStop}%
\bibitem [{\citenamefont {Lhuillier}\ and\ \citenamefont {Misguich}(2002)}]{lhuillier2002frustrated}%
  \BibitemOpen
  \bibfield  {author} {\bibinfo {author} {\bibfnamefont {C.}~\bibnamefont {Lhuillier}}\ and\ \bibinfo {author} {\bibfnamefont {G.}~\bibnamefont {Misguich}},\ }in\ \href@noop {} {\emph {\bibinfo {booktitle} {High Magnetic Fields: Applications in Condensed Matter Physics and Spectroscopy}}}\ (\bibinfo  {publisher} {Springer},\ \bibinfo {year} {2002})\ pp.\ \bibinfo {pages} {161--190}\BibitemShut {NoStop}%
\bibitem [{\citenamefont {Laeuchli}\ \emph {et~al.}(2016)\citenamefont {Laeuchli}, \citenamefont {Schuler},\ and\ \citenamefont {Wietek}}]{laeuchli2016studying}%
  \BibitemOpen
  \bibfield  {author} {\bibinfo {author} {\bibfnamefont {A.}~\bibnamefont {Laeuchli}}, \bibinfo {author} {\bibfnamefont {M.}~\bibnamefont {Schuler}},\ and\ \bibinfo {author} {\bibfnamefont {A.}~\bibnamefont {Wietek}},\ }\href@noop {} {\bibfield  {journal} {\bibinfo  {journal} {Quantum Materials: Experiments and Theory}\ ,\ \bibinfo {pages} {43}} (\bibinfo {year} {2016})}\BibitemShut {NoStop}%
\bibitem [{\citenamefont {Beekman}(2015)}]{beekman2015criteria}%
  \BibitemOpen
  \bibfield  {author} {\bibinfo {author} {\bibfnamefont {A.~J.}\ \bibnamefont {Beekman}},\ }\href@noop {} {\bibfield  {journal} {\bibinfo  {journal} {Annals of Physics}\ }\textbf {\bibinfo {volume} {361}},\ \bibinfo {pages} {461} (\bibinfo {year} {2015})}\BibitemShut {NoStop}%
\bibitem [{\citenamefont {Rademaker}\ \emph {et~al.}(2020)\citenamefont {Rademaker}, \citenamefont {Beekman},\ and\ \citenamefont {van Wezel}}]{rademaker2020stability}%
  \BibitemOpen
  \bibfield  {author} {\bibinfo {author} {\bibfnamefont {L.}~\bibnamefont {Rademaker}}, \bibinfo {author} {\bibfnamefont {A.}~\bibnamefont {Beekman}},\ and\ \bibinfo {author} {\bibfnamefont {J.}~\bibnamefont {van Wezel}},\ }\href {https://doi.org/10.1103/PhysRevResearch.2.013304} {\bibfield  {journal} {\bibinfo  {journal} {Phys. Rev. Res.}\ }\textbf {\bibinfo {volume} {2}},\ \bibinfo {pages} {013304} (\bibinfo {year} {2020})}\BibitemShut {NoStop}%
\bibitem [{\citenamefont {Sandvik}\ and\ \citenamefont {Kurkij\"arvi}(1991)}]{Sandviksusc1991}%
  \BibitemOpen
  \bibfield  {author} {\bibinfo {author} {\bibfnamefont {A.~W.}\ \bibnamefont {Sandvik}}\ and\ \bibinfo {author} {\bibfnamefont {J.}~\bibnamefont {Kurkij\"arvi}},\ }\href {https://doi.org/10.1103/PhysRevB.43.5950} {\bibfield  {journal} {\bibinfo  {journal} {Phys. Rev. B}\ }\textbf {\bibinfo {volume} {43}},\ \bibinfo {pages} {5950} (\bibinfo {year} {1991})}\BibitemShut {NoStop}%
\bibitem [{\citenamefont {Sandvik}(1999)}]{Sandvik1999}%
  \BibitemOpen
  \bibfield  {author} {\bibinfo {author} {\bibfnamefont {A.~W.}\ \bibnamefont {Sandvik}},\ }\href {https://doi.org/10.1103/PhysRevB.59.R14157} {\bibfield  {journal} {\bibinfo  {journal} {Phys. Rev. B}\ }\textbf {\bibinfo {volume} {59}},\ \bibinfo {pages} {R14157} (\bibinfo {year} {1999})}\BibitemShut {NoStop}%
\bibitem [{\citenamefont {Yan}\ \emph {et~al.}(2019)\citenamefont {Yan}, \citenamefont {Wu}, \citenamefont {Liu}, \citenamefont {Sylju\aa{}sen}, \citenamefont {Lou},\ and\ \citenamefont {Chen}}]{yan2019sweeping}%
  \BibitemOpen
  \bibfield  {author} {\bibinfo {author} {\bibfnamefont {Z.}~\bibnamefont {Yan}}, \bibinfo {author} {\bibfnamefont {Y.}~\bibnamefont {Wu}}, \bibinfo {author} {\bibfnamefont {C.}~\bibnamefont {Liu}}, \bibinfo {author} {\bibfnamefont {O.~F.}\ \bibnamefont {Sylju\aa{}sen}}, \bibinfo {author} {\bibfnamefont {J.}~\bibnamefont {Lou}},\ and\ \bibinfo {author} {\bibfnamefont {Y.}~\bibnamefont {Chen}},\ }\href {https://doi.org/10.1103/PhysRevB.99.165135} {\bibfield  {journal} {\bibinfo  {journal} {Phys. Rev. B}\ }\textbf {\bibinfo {volume} {99}},\ \bibinfo {pages} {165135} (\bibinfo {year} {2019})}\BibitemShut {NoStop}%
\bibitem [{\citenamefont {Yan}(2022)}]{yan2022global}%
  \BibitemOpen
  \bibfield  {author} {\bibinfo {author} {\bibfnamefont {Z.}~\bibnamefont {Yan}},\ }\href {https://doi.org/10.1103/PhysRevB.105.184432} {\bibfield  {journal} {\bibinfo  {journal} {Phys. Rev. B}\ }\textbf {\bibinfo {volume} {105}},\ \bibinfo {pages} {184432} (\bibinfo {year} {2022})}\BibitemShut {NoStop}%
\bibitem [{\citenamefont {Binder}(1981)}]{Binder1981}%
  \BibitemOpen
  \bibfield  {author} {\bibinfo {author} {\bibfnamefont {K.}~\bibnamefont {Binder}},\ }\href {https://doi.org/10.1103/PhysRevLett.47.693} {\bibfield  {journal} {\bibinfo  {journal} {Phys. Rev. Lett.}\ }\textbf {\bibinfo {volume} {47}},\ \bibinfo {pages} {693} (\bibinfo {year} {1981})}\BibitemShut {NoStop}%
\bibitem [{\citenamefont {Binder}\ and\ \citenamefont {Landau}(1984)}]{Binder1984}%
  \BibitemOpen
  \bibfield  {author} {\bibinfo {author} {\bibfnamefont {K.}~\bibnamefont {Binder}}\ and\ \bibinfo {author} {\bibfnamefont {D.~P.}\ \bibnamefont {Landau}},\ }\href {https://doi.org/10.1103/PhysRevB.30.1477} {\bibfield  {journal} {\bibinfo  {journal} {Phys. Rev. B}\ }\textbf {\bibinfo {volume} {30}},\ \bibinfo {pages} {1477} (\bibinfo {year} {1984})}\BibitemShut {NoStop}%
\bibitem [{\citenamefont {Bärwinkel}\ \emph {et~al.}(2000)\citenamefont {Bärwinkel}, \citenamefont {Schmidt},\ and\ \citenamefont {Schnack}}]{Schnack2000jmmm}%
  \BibitemOpen
  \bibfield  {author} {\bibinfo {author} {\bibfnamefont {K.}~\bibnamefont {Bärwinkel}}, \bibinfo {author} {\bibfnamefont {H.-J.}\ \bibnamefont {Schmidt}},\ and\ \bibinfo {author} {\bibfnamefont {J.}~\bibnamefont {Schnack}},\ }\href {https://doi.org/https://doi.org/10.1016/S0304-8853(00)00481-9} {\bibfield  {journal} {\bibinfo  {journal} {Journal of Magnetism and Magnetic Materials}\ }\textbf {\bibinfo {volume} {220}},\ \bibinfo {pages} {227} (\bibinfo {year} {2000})}\BibitemShut {NoStop}%
\bibitem [{\citenamefont {Wietek}\ \emph {et~al.}(2017)\citenamefont {Wietek}, \citenamefont {Schuler},\ and\ \citenamefont {L{\"a}uchli}}]{wietek2017studying}%
  \BibitemOpen
  \bibfield  {author} {\bibinfo {author} {\bibfnamefont {A.}~\bibnamefont {Wietek}}, \bibinfo {author} {\bibfnamefont {M.}~\bibnamefont {Schuler}},\ and\ \bibinfo {author} {\bibfnamefont {A.~M.}\ \bibnamefont {L{\"a}uchli}},\ }\href@noop {} {\bibfield  {journal} {\bibinfo  {journal} {arXiv preprint arXiv:1704.08622}\ } (\bibinfo {year} {2017})}\BibitemShut {NoStop}%
\bibitem [{\citenamefont {Mao}\ \emph {et~al.}(2025)\citenamefont {Mao}, \citenamefont {Ding}, \citenamefont {Wang}, \citenamefont {Hu},\ and\ \citenamefont {Yan}}]{mao2025sampling}%
  \BibitemOpen
  \bibfield  {author} {\bibinfo {author} {\bibfnamefont {B.-B.}\ \bibnamefont {Mao}}, \bibinfo {author} {\bibfnamefont {Y.-M.}\ \bibnamefont {Ding}}, \bibinfo {author} {\bibfnamefont {Z.}~\bibnamefont {Wang}}, \bibinfo {author} {\bibfnamefont {S.}~\bibnamefont {Hu}},\ and\ \bibinfo {author} {\bibfnamefont {Z.}~\bibnamefont {Yan}},\ }\href {https://doi.org/10.1038/s41467-025-58058-0} {\bibfield  {journal} {\bibinfo  {journal} {Nat. Commun}\ }\textbf {\bibinfo {volume} {16}},\ \bibinfo {pages} {2880} (\bibinfo {year} {2025})}\BibitemShut {NoStop}%
\bibitem [{\citenamefont {Mao}\ \emph {et~al.}(2026)\citenamefont {Mao}, \citenamefont {Wang}, \citenamefont {Chen},\ and\ \citenamefont {Yan}}]{mao2025arxiv}%
  \BibitemOpen
  \bibfield  {author} {\bibinfo {author} {\bibfnamefont {B.-B.}\ \bibnamefont {Mao}}, \bibinfo {author} {\bibfnamefont {Z.}~\bibnamefont {Wang}}, \bibinfo {author} {\bibfnamefont {B.-B.}\ \bibnamefont {Chen}},\ and\ \bibinfo {author} {\bibfnamefont {Z.}~\bibnamefont {Yan}},\ }\href {https://doi.org/10.1103/7j21-l3pg} {\bibfield  {journal} {\bibinfo  {journal} {Phys. Rev. Lett.}\ }\textbf {\bibinfo {volume} {136}},\ \bibinfo {pages} {046401} (\bibinfo {year} {2026})}\BibitemShut {NoStop}%
\bibitem [{\citenamefont {Sandvik}(2016)}]{sandvik2016constrained}%
  \BibitemOpen
  \bibfield  {author} {\bibinfo {author} {\bibfnamefont {A.~W.}\ \bibnamefont {Sandvik}},\ }\href {https://doi.org/10.1103/PhysRevE.94.063308} {\bibfield  {journal} {\bibinfo  {journal} {Phys. Rev. E}\ }\textbf {\bibinfo {volume} {94}},\ \bibinfo {pages} {063308} (\bibinfo {year} {2016})}\BibitemShut {NoStop}%
\bibitem [{\citenamefont {Shao}\ \emph {et~al.}(2017)\citenamefont {Shao}, \citenamefont {Qin}, \citenamefont {Capponi}, \citenamefont {Chesi}, \citenamefont {Meng},\ and\ \citenamefont {Sandvik}}]{shao2017nearly}%
  \BibitemOpen
  \bibfield  {author} {\bibinfo {author} {\bibfnamefont {H.}~\bibnamefont {Shao}}, \bibinfo {author} {\bibfnamefont {Y.~Q.}\ \bibnamefont {Qin}}, \bibinfo {author} {\bibfnamefont {S.}~\bibnamefont {Capponi}}, \bibinfo {author} {\bibfnamefont {S.}~\bibnamefont {Chesi}}, \bibinfo {author} {\bibfnamefont {Z.~Y.}\ \bibnamefont {Meng}},\ and\ \bibinfo {author} {\bibfnamefont {A.~W.}\ \bibnamefont {Sandvik}},\ }\href {https://doi.org/10.1103/PhysRevX.7.041072} {\bibfield  {journal} {\bibinfo  {journal} {Phys. Rev. X}\ }\textbf {\bibinfo {volume} {7}},\ \bibinfo {pages} {041072} (\bibinfo {year} {2017})}\BibitemShut {NoStop}%
\bibitem [{\citenamefont {Shao}\ and\ \citenamefont {Sandvik}(2023)}]{shao2023progress}%
  \BibitemOpen
  \bibfield  {author} {\bibinfo {author} {\bibfnamefont {H.}~\bibnamefont {Shao}}\ and\ \bibinfo {author} {\bibfnamefont {A.~W.}\ \bibnamefont {Sandvik}},\ }\href {https://doi.org/https://doi.org/10.1016/j.physrep.2022.11.002} {\bibfield  {journal} {\bibinfo  {journal} {Physics Reports}\ }\textbf {\bibinfo {volume} {1003}},\ \bibinfo {pages} {1} (\bibinfo {year} {2023})}\BibitemShut {NoStop}%
\bibitem [{\citenamefont {Wu}\ \emph {et~al.}(1999)\citenamefont {Wu}, \citenamefont {Chen}, \citenamefont {Dai}, \citenamefont {Yu},\ and\ \citenamefont {Su}}]{PhysRevB.60.1057}%
  \BibitemOpen
  \bibfield  {author} {\bibinfo {author} {\bibfnamefont {C.}~\bibnamefont {Wu}}, \bibinfo {author} {\bibfnamefont {B.}~\bibnamefont {Chen}}, \bibinfo {author} {\bibfnamefont {X.}~\bibnamefont {Dai}}, \bibinfo {author} {\bibfnamefont {Y.}~\bibnamefont {Yu}},\ and\ \bibinfo {author} {\bibfnamefont {Z.-B.}\ \bibnamefont {Su}},\ }\href {https://doi.org/10.1103/PhysRevB.60.1057} {\bibfield  {journal} {\bibinfo  {journal} {Phys. Rev. B}\ }\textbf {\bibinfo {volume} {60}},\ \bibinfo {pages} {1057} (\bibinfo {year} {1999})}\BibitemShut {NoStop}%
\bibitem [{\citenamefont {Langari}(1998)}]{Langari1998prb}%
  \BibitemOpen
  \bibfield  {author} {\bibinfo {author} {\bibfnamefont {A.}~\bibnamefont {Langari}},\ }\href {https://doi.org/10.1103/PhysRevB.58.14467} {\bibfield  {journal} {\bibinfo  {journal} {Phys. Rev. B}\ }\textbf {\bibinfo {volume} {58}},\ \bibinfo {pages} {14467} (\bibinfo {year} {1998})}\BibitemShut {NoStop}%
\bibitem [{\citenamefont {Kargarian}\ \emph {et~al.}(2008)\citenamefont {Kargarian}, \citenamefont {Jafari},\ and\ \citenamefont {Langari}}]{Langari2008pra}%
  \BibitemOpen
  \bibfield  {author} {\bibinfo {author} {\bibfnamefont {M.}~\bibnamefont {Kargarian}}, \bibinfo {author} {\bibfnamefont {R.}~\bibnamefont {Jafari}},\ and\ \bibinfo {author} {\bibfnamefont {A.}~\bibnamefont {Langari}},\ }\href {https://doi.org/10.1103/PhysRevA.77.032346} {\bibfield  {journal} {\bibinfo  {journal} {Phys. Rev. A}\ }\textbf {\bibinfo {volume} {77}},\ \bibinfo {pages} {032346} (\bibinfo {year} {2008})}\BibitemShut {NoStop}%
\bibitem [{\citenamefont {Karbach}\ and\ \citenamefont {M{\"u}ller}(1997)}]{karbach1997}%
  \BibitemOpen
  \bibfield  {author} {\bibinfo {author} {\bibfnamefont {M.}~\bibnamefont {Karbach}}\ and\ \bibinfo {author} {\bibfnamefont {G.}~\bibnamefont {M{\"u}ller}},\ }\href@noop {} {\bibfield  {journal} {\bibinfo  {journal} {arXiv preprint cond-mat/9809162}\ } (\bibinfo {year} {1997})}\BibitemShut {NoStop}%
\bibitem [{\citenamefont {Nahum}(2025)}]{Nahum2025}%
  \BibitemOpen
  \bibfield  {author} {\bibinfo {author} {\bibfnamefont {A.}~\bibnamefont {Nahum}},\ }\href@noop {} {\bibfield  {journal} {\bibinfo  {journal} {arXiv preprint arXiv:2506.21540}\ } (\bibinfo {year} {2025})}\BibitemShut {NoStop}%
\bibitem [{\citenamefont {Wang}(2026)}]{wang}%
  \BibitemOpen
  \bibfield  {author} {\bibinfo {author} {\bibfnamefont {J.-Y.}\ \bibnamefont {Wang}},\ }\bibfield  {journal} {\bibinfo  {journal} {Zenodo}\ }\href {https://doi.org/10.5281/zenodo.19565431} {10.5281/zenodo.19565431} (\bibinfo {year} {2026})\BibitemShut {NoStop}%
\bibitem [{\citenamefont {Liu}\ \emph {et~al.}(2022)\citenamefont {Liu}, \citenamefont {Li}, \citenamefont {Huang}, \citenamefont {Li}, \citenamefont {Yan},\ and\ \citenamefont {Yao}}]{liu2022bulk}%
  \BibitemOpen
  \bibfield  {author} {\bibinfo {author} {\bibfnamefont {Z.}~\bibnamefont {Liu}}, \bibinfo {author} {\bibfnamefont {J.}~\bibnamefont {Li}}, \bibinfo {author} {\bibfnamefont {R.-Z.}\ \bibnamefont {Huang}}, \bibinfo {author} {\bibfnamefont {J.}~\bibnamefont {Li}}, \bibinfo {author} {\bibfnamefont {Z.}~\bibnamefont {Yan}},\ and\ \bibinfo {author} {\bibfnamefont {D.-X.}\ \bibnamefont {Yao}},\ }\href {https://doi.org/10.1103/PhysRevB.105.014418} {\bibfield  {journal} {\bibinfo  {journal} {Phys. Rev. B}\ }\textbf {\bibinfo {volume} {105}},\ \bibinfo {pages} {014418} (\bibinfo {year} {2022})}\BibitemShut {NoStop}%
\bibitem [{\citenamefont {Toth}\ and\ \citenamefont {Lake}(2015)}]{Toth2015}%
  \BibitemOpen
  \bibfield  {author} {\bibinfo {author} {\bibfnamefont {S.}~\bibnamefont {Toth}}\ and\ \bibinfo {author} {\bibfnamefont {B.}~\bibnamefont {Lake}},\ }\href {https://doi.org/10.1088/0953-8984/27/16/166002} {\bibfield  {journal} {\bibinfo  {journal} {Journal of Physics: Condensed Matter}\ }\textbf {\bibinfo {volume} {27}},\ \bibinfo {pages} {166002} (\bibinfo {year} {2015})}\BibitemShut {NoStop}%
\bibitem [{\citenamefont {Xiong}\ \emph {et~al.}(2020)\citenamefont {Xiong}, \citenamefont {Datta},\ and\ \citenamefont {Yao}}]{Xiong2020npj}%
  \BibitemOpen
  \bibfield  {author} {\bibinfo {author} {\bibfnamefont {Z.}~\bibnamefont {Xiong}}, \bibinfo {author} {\bibfnamefont {T.}~\bibnamefont {Datta}},\ and\ \bibinfo {author} {\bibfnamefont {D.-X.}\ \bibnamefont {Yao}},\ }\href {https://doi.org/10.1038/s41535-020-00282-6} {\bibfield  {journal} {\bibinfo  {journal} {npj Quantum Materials}\ }\textbf {\bibinfo {volume} {5}},\ \bibinfo {pages} {78} (\bibinfo {year} {2020})}\BibitemShut {NoStop}%
\bibitem [{\citenamefont {Gao}\ and\ \citenamefont {Wu}(2025)}]{cpl_42_4_047501}%
  \BibitemOpen
  \bibfield  {author} {\bibinfo {author} {\bibfnamefont {Y.}~\bibnamefont {Gao}}\ and\ \bibinfo {author} {\bibfnamefont {J.}~\bibnamefont {Wu}},\ }\href {https://doi.org/10.1088/0256-307X/42/4/047501} {\bibfield  {journal} {\bibinfo  {journal} {Chin. Phys. Lett.}\ }\textbf {\bibinfo {volume} {42}},\ \bibinfo {pages} {047501} (\bibinfo {year} {2025})}\BibitemShut {NoStop}%
\end{thebibliography}%

\begin{widetext}

\appendix

\newpage

\section{Demonstrating the system is Non-frustration-free}

For a Hamiltonian that can be decomposed into a sum of local terms, i.e., $\hat{H} = \sum_{i=1}^N \hat{h}_i$, if its ground state $|\psi\rangle$ is also the ground state of each local term $\hat{h}_i$ (satisfying $\hat{h}_i |\psi\rangle = \varepsilon_i |\psi\rangle$), and the ground-state energy of the whole system equals the sum of the ground state energies of the local terms [i.e., $\hat{H} |\psi\rangle = (\sum_{i=1}^N \varepsilon_i) |\psi\rangle$], then such a Hamiltonian is called a frustration-free Hamiltonian.

To verify whether the Hamiltonian of the system is a frustration-free Hamiltonian, we conducted the following analysis: First, we calculated the ground-state energy ($\varepsilon$) and ground state ($|\psi\rangle $) of the Hamiltonian for systems of different sizes ($L=4, 6, 8$), as well as the ground-state energy ($  \varepsilon_i$) and ground state ($|\varphi_i  \rangle $) of each local term $\hat{h}_i=V\textbf{S}_{i,A}\cdot\textbf{S}_{i,C}+V_{1}\textbf{S}_{i,A}\cdot\textbf{S}_{i,B}+V_1\textbf{S}_{i,C}\cdot \textbf{S}_{i+1,A}$ that constitutes the Hamiltonian. We set $V=V_1=1$. By computing the inner product $\langle \psi|\varphi_i  \rangle$ between the ground state of the system and the ground state of each local term, we found that the results are all close to zero (as shown in Table \ref{tab:data}), indicating that the ground state of the system is not the ground state of any local term. In addition, the ground-state energy of the system $\varepsilon$ is not equal to the sum of the ground state energies of the local terms $\sum_{i=1}^N \varepsilon_i$. These two results together demonstrate that the Hamiltonian does not satisfy the defining conditions of a frustration-free Hamiltonian.

\begin{table}[htbp]
  \centering
  \caption{Ground state inner product and energies of $\hat{H}$ and local terms $\hat{h}_i$ in systems of different sizes}  
  \label{tab:data}
  \begin{tabular}{cccccccccccc}  
    \toprule
    \midrule
    L & $\langle\psi|\varphi_1\rangle$ & $\langle\psi|\varphi_2\rangle$ & $\langle\psi|\varphi_3\rangle$ & $\langle\psi|\varphi_4\rangle$
    &$\varepsilon$ &$\varepsilon_1$ &$\varepsilon_2$ &$\varepsilon_3$ &$\varepsilon_4$ &$\sum_i\varepsilon_i$ & $\varepsilon-\sum_i\varepsilon_i$\\
    4   & -0.0175   & 0.00986 & \empty & \empty &-2.45 & -1.62 & -1.62 & \empty & \empty &-3.23& 0.78 \\
    6   & 0.0967   &0.00870  & -0.156  & \empty &-3.61  & -1.62 &-1.62 &-1.62 & \empty & -4.85& 1.24\\
    8   & -0.0401   & -0.00612 &-0.00233   & 0.00321 & -5.28 &-1.62 &-1.62 &-1.62 &-1.62 &-6.46& 1.18\\
    \bottomrule
  \end{tabular}
\end{table}

\section{Energy spectrum is robust}
\label{appendixb}
\begin{figure}[htbp]
	\centering
	\includegraphics[width=0.9\textwidth]{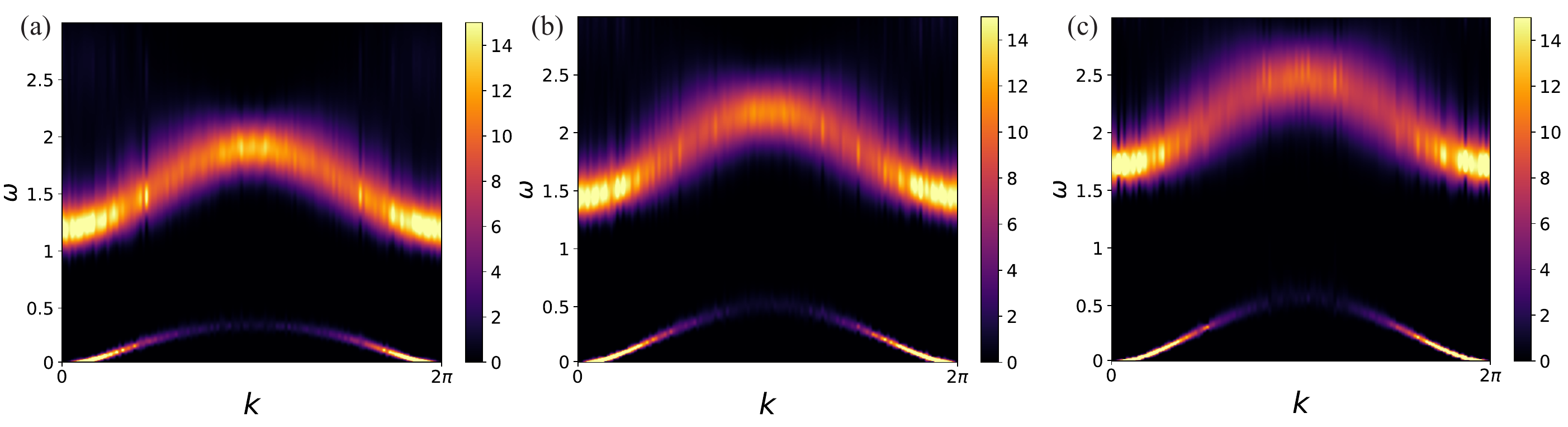}
	\caption{\label{xx} Spectrum calculated by QMC-SAC for $L = 256$, where $V = 1$, (a) $V_1 = 0.8$, (b)$V_1 = 1$ and (c) $V_1 = 1.2$.}
    \label{sacsm}
\end{figure}

We can compute the spin correlation functions between different sublattices (A, B, C) within unit cells (i, j) using the SSE method
\begin{equation}
c_{ij}^k(r_{ij}, \tau) = \langle s_{\alpha i}^z(\tau) s_{\beta j}^z \rangle,
\end{equation}
where ( i, j ) label different unit cells,  $( \alpha, \beta \in {A, B, C} )$ denote the sublattice positions, and $ c_{ij}^k(r_{ij}, \tau)$ represents the correlation between the spin on sublattice $\alpha$ in unit cell $i$ and the spin on sublattice $\beta$ in unit cell $j$ at imaginary time $\tau$ and momentum point $k$.

By traversing all intercell distances $r_{ij}$ and performing a Fourier transform on the spin correlations between sublattices, we obtain the imaginary-time correlation function in momentum space
\begin{equation}
c_{\alpha\beta}^k(\tau) = \sum_{r_{ij}=0}^{L/2 - 1} e^{-i k r_{ij}} c_{ij}^k(r_{ij}, \tau).
\end{equation}

This allows us to construct the correlation matrix in momentum space:
\begin{equation}
\begin{pmatrix}
C_{AA}^k(\tau) & C_{AB}^k(\tau) & C_{AC}^k(\tau) \\
C_{BA}^k(\tau) & C_{BB}^k(\tau) & C_{BC}^k(\tau) \\
C_{CA}^k(\tau) & C_{CB}^k(\tau) & C_{CC}^k(\tau)
\end{pmatrix}.
\end{equation}

By diagonalizing this matrix and summing over its three eigenvalues, we obtain the imaginary-time of the full system correlation function at momentum $k$, denoted $C(k, \tau)$.

According to the definition of the spectral function, the imaginary-time correlation function $C(k, \tau)$ is related to the real frequency spectral function $S(\omega)$ via a Laplace transform
\begin{equation}
C(k, \tau) = \int_0^\infty \frac{e^{-\tau \omega} + e^{-(\beta - \tau) \omega}}{\pi} S(\omega) \, d\omega,
\end{equation}

Using the SAC method \cite{sandvik2016constrained, shao2017nearly, shao2023progress}, we can reconstruct the excitation spectrum of the system from the imaginary-time correlation function. 

Specifically, the uncertainty of this method mainly comes from two sources. The first is the statistical error of the imaginary-time correlation function: Our data are obtained by averaging over 60 independent bins, with each bin run on a different core and using a different random seed, leading to an error of order $10^{-5}$. The second is the systematic uncertainty introduced by the SAC procedure itself. However, the spectral function presented in the manuscript is the average over SAC sampling results. 

To examine the reproducibility of the spectrum and the robustness of the SAC procedure, we take the case with $L=256$ and $V=1$  and compute the imaginary-time correlation functions for $V_1 = 0.8, 1.0$, and $1.2$, followed by spectral reconstruction using the SAC. As shown in Fig.~\ref{xx}, although the peak intensity increases with increasing $V_1$, the main spectral feature relevant to our conclusion—particularly the low-energy quadratic dispersion near $k\sim 0$ —remains stable for all three parameter sets. In addition, Fig.~\ref{rob,spinwavedis} presents the results obtained from linear spin-wave theory and from the low-energy effective model constructed using Kadanoff’s approach. Both show a quadratic dispersion near $k \sim 0$, consistent with the results obtained from SSE+SAC. The SAC method has by now become relatively mature and has been widely used in calculations of spectral functions in quantum many-body systems. For example, Refs.~\cite{sandvik2016constrained, shao2017nearly, liu2022bulk} all employed this method to obtain reliable dynamical information.

\section{Proof of Robustness }

\begin{figure}[htbp]
	\centering
	\includegraphics[width=0.6\textwidth]{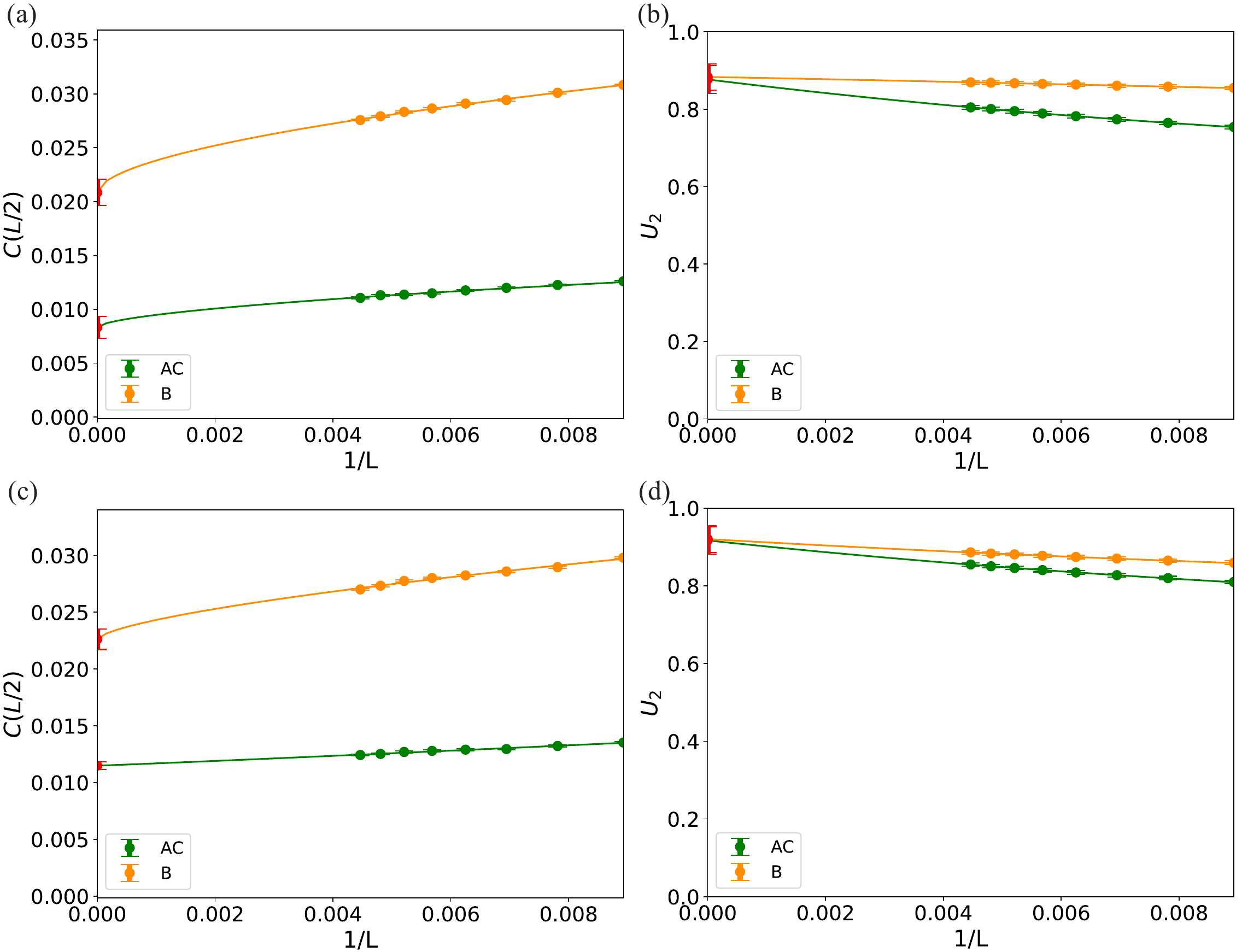}
	\caption{\label{cu} Correlation function $C(L/2)$ and Binder cumulant $ U_2$ vs the inverse of the system size $1/L$ for (a,b) $V = 1, V_1 = 0.8$ and (c,d) $V = 1, V_1 = 1.2$. }
\end{figure}

\begin{equation}
U_{2}(L) =U_{2}+c_{1}L^{-1}+c_{2}L^{-2}+c_{3}L^{-3}.
\label{cs1psm}
\end{equation}

\begin{equation}
    C(L/2) = c + aL^{-p}.
    \label{gapless_surfacesm}
\end{equation}

We selected two representative parameter points $V_1 = 0.8$ and $V_1 = 1.2$ to calculate the Binder cumulant $U_2$ for sublattice $B$  and $AC$. As shown in Fig.~\ref{cu} (b) and (d), the dependence of $U_2(L)$ on $1/L$ can be well fitted by a polynomial in $1/L$ as described by Eq.~(\ref{cs1psm}). The extrapolated results in the thermodynamic limit ($1/L \to 0$) are as follows: for $V_1 = 0.8$, the estimated Binder cumulant is $U_2 = 0.88(3)$ for sublattice $B$ and $U_2 = 0.88(4)$ for sublattice $AC$, both of which are close to 1 within error bars, and for $V_1 = 1.2$, the estimated Binder cumulant is $U_2 = 0.92(3)$ for sublattice $B$ and $U_2 = 0.92(4)$ for sublattice $AC$, both of which are close to 1 within error bars. 

 Under the same parameter conditions, we calculated the correlation function $C(L/2)$ for the spin pairs with the maximum spatial distance $|i - j| = L/2$ at sublattice $B$ and $AC$.
As shown in Fig.~\ref{cu} (a) and (c), both types of correlation functions converge to finite values as the system size $L $ increases. Fitting by Eq.~(\ref{gapless_surfacesm}), we extrapolated the results to the thermodynamic limit ($L \to \infty$). For $V_1 = 0.8$, the extrapolated values are $c = 0.020(1)$ for sublattice $B$ and $c = 0.008(1)$ for sublattice $AC$; for $V_1 = 1.2$, the  corresponding values are $c = 0.0226(9)$ and $0.0115(3)$, respectively. These results further confirm the stability of the symmetry breaking, and the stable quadratic dispersion observed under different parameters also supports this conclusion (see  Fig.~\ref{sacsm}).

\section{Detail of linear spin wave theory}
\label{appendixd}
Because the long-range order is the N\'eel staggered order, first, we define a global frame for spin denoted by $x^0,y^0,z^0$, and the sublattice-dependent spin-quantization axes are considered in the local coordinate system $x,y,z$. The operators in two coordinate systems are related by 
\begin{equation}
\begin{aligned}
S^{x^0}_{A}=&S^{x}_{A},\,S^{y^0}_{A}=S^{y}_{A},\,S^{z^0}_{A}=S^{z}_{A},\\
S^{x^0}_{B}=&-S^{x}_{B},\,S^{y^0}_{B}=S^{y}_{B},\,S^{z^0}_{B}=-S^{z}_{B},\\
S^{x^0}_{C}=&-S^{x}_{C},\,S^{y^0}_{C}=S^{y}_{C},\,S^{z^0}_{C}=-S^{z}_{C}.
\end{aligned}
\end{equation}

In the local coordinate system, we can use the Holstein-Primakoff transformation and linear spin wave approximation, then the spin operators can be expressed by bosonic operators
\begin{equation}
S^{z}_{i,\alpha}=S-a_{i,\alpha}^{+}a_{i,\alpha},
S^{+}_{i,\alpha}\approx\sqrt{2S}a_{i,\alpha},
S^{-}_{i,\alpha}\approx\sqrt{2S}a_{i,\alpha}^{+},
\end{equation}
where $\alpha=A,B,C$ represents the sublattices, and $S^{+}=S^{x}+iS^{y}$ and $S^{-}=S^{x}-iS^{y}$. Fourier transformation is defined by
$a_{i,\alpha}=\frac{1}{\sqrt{N}}\sum_{k}a_{\alpha,k}e^{ik(R_{i}+r_{\alpha})}$,
where $N$ is the unit cell number, $R_i$ is the position of the $i$th unitcell, and $r_\alpha$ is the position of $\alpha$ site in $i$th unitcell. The lattice constant is the distance between two nearest-neighbor $A$ sites and it is set to be 1. Here, $r_{A}=0=r_{B}, r_{C}=1/2$ for the comb chain. After the Fourier transformation, we find the linear spin-wave Hamiltonian
\begin{equation}
\begin{aligned}
H^{LSW} = &\frac{1}{2}\sum_{k}[-(Ve^{-ik/2}+V_1e^{ik/2})a_{A,k}a_{C,-k}-(Ve^{ik/2}+V_1e^{-ik/2})a^{+}_{A,k}a^{+}_{C,-k}-V_1a_{A,k}a_{B,-k}-V_1a^{+}_{A,k}a^{+}_{B,-k}\\
&+(V+V_1)a^{+}_{C,k}a_{C,k}+(V+2V_1)a^{+}_{A,k}a_{A,k}+V_1a^{+}_{B,k}a_{B,k}].
\end{aligned}
\end{equation}
The spin wave dispersions can be obtained by diagonalizing this quadratic linear spin wave Hamiltonian via Bogoliubov transformation \cite{Toth2015,Xiong2020npj,cpl_42_4_047501}.

\section{Real-space renormalization group to obtain the low-energy effective model}
\label{appendixe}
In this section, we will show that the low-energy effective model here is a ferromagnetic Heisenberg chain with quadratic dispersion.
According to Kadanoff's approach \cite{Langari1998prb,Langari2008pra}, the model can be divided into two parts, intra block $H^B$ and inter-block $H^{BB}$. The intra block Hamiltonian is the sum of individual blocks. Then, each block can be treated exactly to find the low-energy states and construct the projection operator $P$ onto the low energy subspace. The effective Hamiltonian can be obtained by projecting the inter-block part onto the low energy subspace $H^{eff}=PH^{BB}P^{\dagger}$. We choose a three-site block; then $H^B$ can be written as
\begin{equation}
\begin{aligned}
H^B=&\sum_{I}h^{B}_I\\
=&\sum_{I}V_1\left(\textbf{S}_{I,C}\cdot \textbf{S}_{I,A}+\textbf{S}_{I,A}\cdot\textbf{S}_{I,B}\right),
\end{aligned}
\end{equation}
where, $I$ labels the block. 

First, we want to find the ground states of $h^{B}_I$. It is obvious that a symmetry subgroup of the Hamiltonian is U(1), namely, rotation around the $z$ axis. Then we can use the quantum number $M^z$ to block diagonalize the $h^{B}_I$, and there is spin-flip symmetry $U^{\alpha}_{\pi}$ ($\pi$ rotation around the $\alpha$, $\alpha=x,y$), thus, if $|\psi\rangle$ is a ground state with $M^z$, then $U^{\alpha}_{\pi}|\psi\rangle$ is also a ground state with $-M^z$. 

For this three-site block, $M^z=-3/2,-1/2,1/2,3/2$, it is not difficult to find that the ground state should be in the $M^z=\pm 1/2$ sectors. The energy of the ground state of $h^{B}_I$ is $-V_1$. The ground state wave function in $M^{z}=1/2$ sector is
\begin{equation}
|\psi_0\rangle=\frac{1}{\sqrt{6}}\left(|\uparrow\uparrow\downarrow\rangle-2|\uparrow\downarrow\uparrow\rangle+|\downarrow\uparrow\uparrow\rangle\right)\equiv|\Uparrow\rangle,
\end{equation}
and the ground state wave function in $M^{z}=-1/2$ sector is 
\begin{equation}
|\psi'_0\rangle=\frac{1}{\sqrt{6}}\left(|\downarrow\downarrow\uparrow\rangle-2|\downarrow\uparrow\downarrow\rangle+|\uparrow\downarrow\downarrow\rangle\right)\equiv|\Downarrow\rangle.
\end{equation}

Secondly, we want to find the low energy operator. The projection operator for the $I$th block is defined by
\begin{equation}
P_{I}=|\Uparrow\rangle_{I}\, _I\langle \psi_0|+|\Downarrow\rangle_{I}\,_I\langle \psi'_0|,
\end{equation}
the full projection operator is simply 
\begin{equation}
P=\prod_{I}P_I.
\end{equation}
In our choice of $h^{B}_I$, the corresponding inter-block Hamiltonian is given by
\begin{equation}
H^{BB}=V\sum_{I}\textbf{S}_{I,A}\cdot \textbf{S}_{I+1,C}.
\end{equation}
thus, we just need to work out the low energy operator for $S_{A}$ and $S_C$. We find that 
\begin{equation}
\begin{aligned}
P_{I}\sigma^{x}_{A}P_{I}^{\dagger}=&\frac{1}{3}\tilde{\sigma}^{x},~~~~~
P_{I}\sigma^{y}_{A}P_{I}^{\dagger}=\frac{1}{3}\tilde{\sigma}^{y},\\
P_{I}\sigma^{z}_{A}P_{I}^{\dagger}=&-\frac{1}{3}\tilde{\sigma}^{z},~
P_{I}\sigma^{x}_{C}P_{I}^{\dagger}=-\frac{2}{3}\tilde{\sigma}^{x},\\
P_{I}\sigma^{y}_{C}P_{I}^{\dagger}=&-\frac{2}{3}\tilde{\sigma}^{y},~
P_{I}\sigma^{z}_{C}P_{I}^{\dagger}=\frac{2}{3}\tilde{\sigma}^{z}.
\end{aligned}
\end{equation}
where the low-energy effective spin-1/2 Pauli operator $\tilde{\sigma}$ is defined by
$\tilde{\sigma}^{x}=|\Uparrow\rangle\langle\Downarrow|+|\Downarrow\rangle\langle\Uparrow|,\
\tilde{\sigma}^{y}=i|\Downarrow\rangle\langle\Uparrow|-i|\Uparrow\rangle\langle\Downarrow|,\
\tilde{\sigma}^{z}=|\Uparrow\rangle\langle\Uparrow|-|\Downarrow\rangle\langle\Downarrow|.
$

Finally, the low-energy effective Hamiltonian is
\begin{equation}
H^{eff}=PH^{BB}P^{\dagger}=-\frac{V}{18}[\sum_{I}\tilde{\sigma}^{x}_{I}\tilde{\sigma}^{x}_{I+1}+\tilde{\sigma}^{y}_{I}\tilde{\sigma}^{y}_{I+1}+\tilde{\sigma}^{z}_{I}\tilde{\sigma}^{z}_{I+1}]=-\frac{2V}{9}\sum_{I}\tilde{S}_{I}\cdot \tilde{S}_{I+1},
\end{equation}
which is a ferromagnetic spin-1/2 chain. Hence, we obtain the well-known quadratic dispersion $2SJ_{eff}(1-\cos k)$ in low energy \cite{karbach1997}, as shown in Fig.~\ref{rob,spinwavedis} (b). The dashed line matches well with the numerical result of QMC-SAC, which further demonstrates the correctness of the low-energy effective Hamiltonian.

It should be noted that the choice of the block is not unique; we can also choose $h^{B}_I=V_{1}\textbf{S}_{I,B}\cdot\textbf{S}_{I,A}+V\textbf{S}_{I,A}\cdot\textbf{S}_{I,C}$. We have confirmed that these two choices give the consistent results, i.e., another choice of $h^{B}_I$ also leads to a ferromagnetic spin-1/2 chain after the projection onto the low-energy subspace.

\section{Bogoliubov inequality and HMW theorem for understanding SSB}
\label{appendixf}
Bogoliubov inequality is a more essential condition that allows for long-range ordered existence~\cite{Watanabe2024Critical}. Here, we verify whether this system conforms to this condition.
The zero temperature version of Bogoliubov inequality is given in ref. \cite{Watanabe2024Critical}
\begin{equation}
\frac{1}{V^2}\sum_k\langle \hat{X}^{\dagger}_{k}\hat{X}_{k}+\hat{X}_{k}\hat{X}^{\dagger}_{k}\rangle\geq\frac{1}{V^2}\sum_{k}\frac{\omega_{k}(h)\left|\langle[i\hat{Q}^{\dagger}_k,\,\hat{X}_{k}]\rangle\right|^2}{\langle[\hat{Q}_k,\,[\hat{H}(h),\,\hat{Q}^{\dagger}_{k}]]\rangle}, 
\end{equation}
where $\hat{H}(h)=H-h\mathcal{O}$, $\mathcal{O}$ is an order parameter, $Q_k$ is the symmetry generator $Q=\sum_i Q_i$ after Fourier transformation, i.e., $Q_k=\sum_i Q_i e^{ik\cdot r_i}$, and $\omega_k(h)$ is the lowest excitation energy of the momentum $k$ sector (we have assumed the translation symmetry implicitly in this inequality). Here, $\hat{X}$ is a Hermitian operator. 

Now, we consider the denominator
\begin{equation}
\begin{aligned}
&\lim_{V\to\infty}\frac{1}{V}\langle [\hat{Q}_{k},[\hat{H}(h),\,\hat{Q}_{k}^{\dagger}]]\rangle\\
=&\lim_{V\to\infty}\frac{1}{V}\langle [\hat{Q}_{k},[\hat{H}-h\hat{O},\,\hat{Q}_{k}^{\dagger}]]\rangle\\
=&\lim_{V\to\infty}\frac{1}{V}\langle [\hat{Q}_{k},[\hat{H},\,\hat{Q}_{k}^{\dagger}]-h[\hat{O},\,\hat{Q}_{k}^{\dagger}]]\rangle\\
=&\lim_{V\to\infty}\frac{1}{V}\langle [\hat{Q}_{k},[\hat{H},\,\hat{Q}_{k}^{\dagger}]]-[\hat{Q}_{k}, h[\hat{O},\,\hat{Q}_{k}^{\dagger}]]\rangle\\
=&\lim_{V\to\infty}\frac{1}{V}\{\langle [\hat{Q}_{k},[\hat{H},\,\hat{Q}_{k}^{\dagger}]]\rangle-\langle[\hat{Q}_{k}, h[\hat{O},\,\hat{Q}_{k}^{\dagger}]]\rangle\}\\
=&A_k+h\,B_k,
\end{aligned}
\end{equation}

Using the Fourier transformation and considering the first term, we obtain
\begin{equation}
\begin{aligned}
A_k=&-\lim_{V\to\infty}\frac{-1}{V}\sum_{i,j}\langle [\hat{Q}_i,[\hat{H},\,\hat{Q}_{j}]]\rangle\cos[k (r_i-r_j)]\\
=&\lim_{V\to\infty}\frac{-1}{V}\sum_{i,j}\langle [\hat{Q}_i,[\hat{H},\,\hat{Q}_{j}]]\rangle
-\lim_{V\to\infty}\frac{-1}{V}\sum_{i,j}\langle [\hat{Q}_i,[\hat{H},\,\hat{Q}_{j}]]\rangle\cos[k (r_i-r_j)]\\
=&\lim_{V\to\infty}\frac{-1}{V}\sum_{i,j}(1-\cos[k (r_i-r_j)])\langle [\hat{Q}_i,[\hat{H},\,\hat{Q}_{j}]]\rangle
\end{aligned}
\end{equation}
where we use the fact that $\sum_{i,j}\langle [\hat{Q}_i,[\hat{H},\,\hat{Q}_{j}]]\rangle=0$ in the second line.
Thus, in the long wave length limit, $A_k\sim k^{2n_0}$ with $n_0=1$. Further, we assume that the lowest excitation energy behaves as $\omega_k(0)\sim |k|^{n}$. Converting the summation into integral \cite{Watanabe2024Critical} and using simple power counting (which gives $k^{n+d-2n_0}$ in the right hand side of Bogoliubov inequality), we have a inequality from Bogoliubov inequality
\begin{equation}
\begin{aligned}
d>2n_0-n,
\end{aligned}
\end{equation}

For the AFM long range order, a widely used order parameter is staggered magnetization $\mathcal{O}=\sum_{i}(-1)^{i}S^{z}_{i}$, where $i$ denotes sites. In this case, continuous symmetry generated by $\hat{Q}=\sum_{i}S^{y}_{i}$ is spontaneously broken, and $\hat{X}$ can be chosen as $\hat{X}=\sum_{i}(-1)^{i}S^{x}_i$, then $\mathcal{O}=[i\hat{Q},\hat{X}]$. 

Additionally, in our comb chain, $d=1, n_0=1, n=2$, thus, $d(=1)>2n_0-n(=2-2)$ which satisfies the inequality, and continuous symmetry breaking is allowed. Normally, for what we commonly refer to as the HMW theorem, $n_0=n=1$, and continuous symmetry breaking is possible only when $d=1$. Similar to frustration-free systems~\cite{Watanabe2024Critical}, our comb chain can bypass the HMW theorem because its excitations are softer than linearly dispersive modes.

\section{Validation of ground-state convergence at $\beta = 2L$}

To  verify whether $\beta = 2L$ is sufficiently low to ensure ground-state convergence, we calculated the convergence of key physical observables—specifically the squared staggered magnetization ($m^2$) and the Binder cumulant ($U_{2}$)—as a function of inverse temperature $\beta$, as shown in Fig.~\ref{beta}. The data  indicate that these quantities converge when $\beta > 1.5L$. We choose $\beta = 2L$. 

 Furthermore, even if $\beta$ were slightly insufficient, it would not alter our main conclusions. Our primary findings concern SSB and the associated long-range order. The presence of long-range order is already unambiguously established at the current $\beta$; increasing $\beta$ further (i.e., lowering the temperature) would only strengthen the signature of this order, not diminish it.
 
\begin{figure}[htbp]
	\centering
	\includegraphics[width=0.7\textwidth]{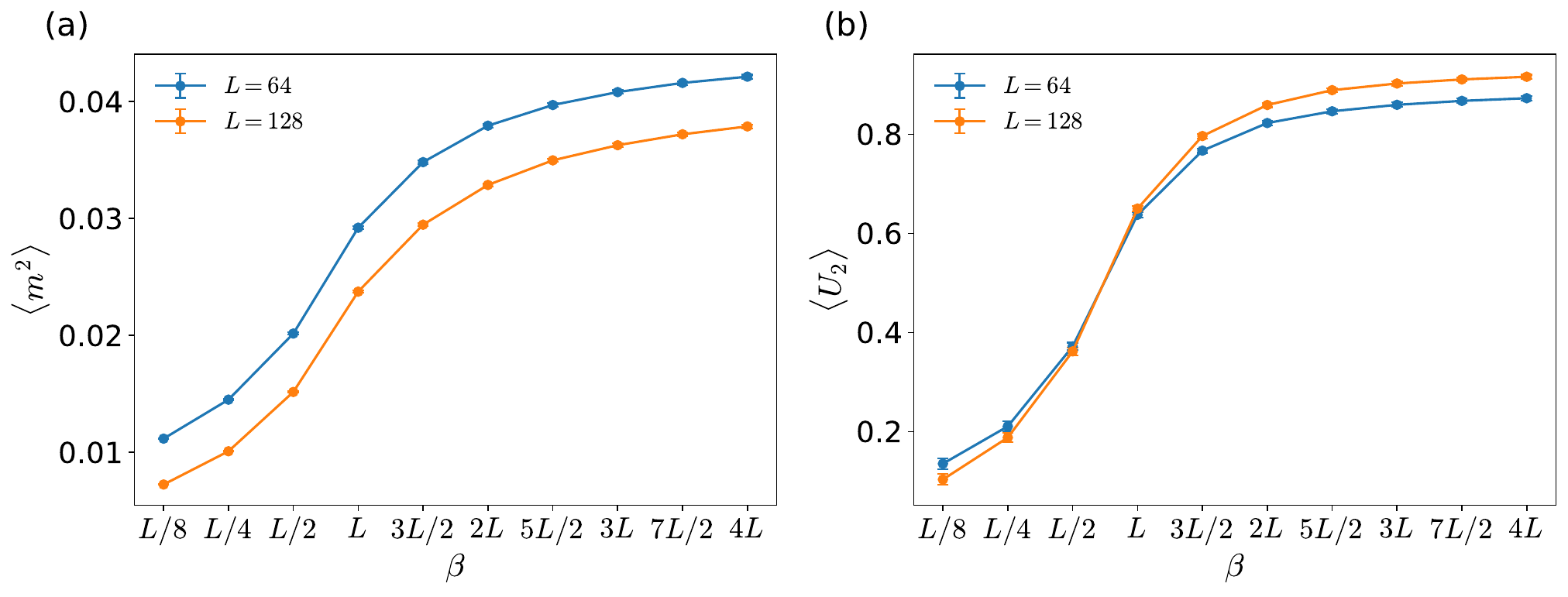}
	\caption{\label{beta} (a) The squared staggered magnetization ($m^2$) and (b) the Binder cumulant ($U_{2}$) as a function of inverse temperature $\beta$ at $V = V_1 = 1$ for the whole system $ABC$. }
\end{figure}
\end{widetext}
\end{document}